\documentclass[12pt]{JHEP3}

\usepackage{amsmath,epsfig}
\usepackage{amssymb,amsfonts}
\usepackage{latexsym}
\usepackage[latin1]{inputenc}
\usepackage{subeqnarray}
\usepackage{xcolor}

\usepackage{graphicx}
\usepackage{longtable}

\relax
\renewcommand{\theequation}{\arabic{section}.\arabic{equation}}
\def\be{\begin{equation}}
\def\ee{\end{equation}}
\def\bea{\begin{eqnarray}}
\def\eea{\end{eqnarray}}

\newcommand\fverb{\setbox\pippobox=\hbox\bgroup\verb}
\newcommand\fverbdo{\egroup\medskip\noindent%
                        \fbox{\unhbox\pippobox}\ }
\newcommand\fverbit{\egroup\item[\fbox{\unhbox\pippobox}]}

\newcommand{\bear}{\begin{eqnarray}}

\newcommand{\eear}{\end{eqnarray}}

\newcommand{\bsea}{\begin{subeqnarray}}
\newcommand{\esea}{\end{subeqnarray}}
\newbox\pippobox

\def\6{\partial}

\def\a{\alpha}
\def\g{\gamma}

\def\pa{\partial}

\def\e{\epsilon}
\def\m{\mu}
\def\n{\nu}
\def\r{\rho}
\def\s{\sigma}
\def\t{\theta}
\def\sp{\;\;\;,\;\;\;}

\def\sq
\def\a{\alpha}

\def\l{\lambda}
\def\y{\psi}

\def\hri#1#2{\href{http://arxiv.org/abs/#1}{[ArXiv:#1]#2}}

\def\e{\epsilon}

\def\d{\delta}

\def\t{\langle J^t\rangle}
\def\x{\langle J^x\rangle}
\def\y{\langle J^y\rangle}
\def\tt{(\langle J^t\rangle-Sh)}
\def\yy{(\langle J^y\rangle+SE)}
\title{\huge Quantum Criticality and DBI Magneto-resistance}
\author{\Large   Elias Kiritsis$^{a,b,c}$, Li Li$^{a,b}$\\
~\\
~\\
$^a$\href{http://hep.physics.uoc.gr}{Crete Center for Theoretical Physics, Institute for Theoretical and Computational Physics},
Department of Physics, University of Crete, 71003 Heraklion, Greece.
~\\
~\\
$^b$Crete Center for Quantum Complexity and Nanotechnology,
Department of Physics, University of Crete, 71003 Heraklion, Greece.
~\\
~\\
$^c$\href{http://www.apc.univ-paris7.fr/APC_CS/}{APC}, Astrparticule et Cosmologie, Universit\'e Paris Diderot, CNRS/IN2P3, CEA/Irfu, Observatoire de Paris, Sorbonne Paris Cit\'e, 10, rue Alice Domon et L\'eonie Duquet, 75205 Paris Cedex 13, France.
~\\
~\\
E-mail: \href{http://hep.physics.uoc.gr/~kiritsis/}{http://hep.physics.uoc.gr/~kiritsis/},
lili@physics.uoc.gr}

\preprint{CCTP-2016-03\\
CCQCN-2016-128}

\abstract{We use the DBI action from string theory and holography to study the magneto-resistance at quantum criticality with hyperscaling violation. We find and analyze a rich class of scaling behaviors for the magneto-resistance. A special case describes the scaling results found in pnictides by  Hayers et al. in~\cite{analytis}. }


\begin{document}

\section{Introduction}
\label{intro}

The holographic duality has been used to understand theories that are at finite density and may be in the universality class of interesting, strongly coupled condensed matter systems. Progress has been made in the application of the holographic description towards phenomena, relevant for condensed
matter systems, in the last few years~\cite{Jan,ammon}. The duality is a natural framework to describe quantum critical systems.

Conductivity is a major observable in condensed matter, which measures the response of a  medium to externally applied electric fields. It has been well known that many correlated electron materials, particularly those thought to be near a quantum critical point (QCP), display a characteristic $T$-linear dependence of the electrical resistivity~\cite{review,norman2011}. Such behavior is distinct from the traditional paradigm of Fermi liquid theory. Many efforts have been made towards understanding  this strange metal behavior, however, a deeper understanding of the $T$-linear resistivity is still an open question.
Several efforts were made to realize such a behavior in holographic materials~\cite{Cu1}-\cite{Amoretti:2016cad}.

The magnetic properties of relevant materials have also proved to be complex and intriguing~\cite{boebinger2009}-\cite{Ong1997}. Several attempts to realize them holographically have been proved unsuccessful. It should be noted however that in the light-cone holographic theory of~\cite{KKP} both resistivity and magneto-resistance were shown to be in agreement with overdoped cuprate behavior at very low temperatures. Holographic scaling techniques were used to investigate the possible realization of both regimes~\cite{kh}. Holographic theories were also shown to provide (under certain conditions) a natural scaling of the AC conductivity~\cite{KPB} not unlike what was seen in cuprates~\cite{VDM}.

More recently, strange metal behavior was observed in the pnictides. Moreover, it has been  shown by experiment that near a putative QCP the magnetic field $h$ plays the same role as the temperature $T$~\cite{analytis}. The measurements imply that the in-plane magneto-resistance behaves as
\begin{equation}\label{linearTh}
\rho_{xx}=\sqrt{\bar{\alpha}\, T^2+\bar{\eta}\, h^2}\,,
\end{equation}
with the magnetic field perpendicular to the plane and $\bar{\alpha},\bar{\eta}$ two constants (see~\cite{analytis} for more details).

Motivated by these results our main goal is to explore holographic theories that reproduce this striking behavior. A natural ingredient in this direction is the DBI action that describes D-branes and which has a suggestive form.

In the fundamental charge case where the U(1) symmetry arises from fundamentals, the gauge field originates from a stack of flavor D-branes. Therefore, the proper gauge field action is the DBI action that includes ``self-interactions" of charge. Computation of the conductivity from a DBI theory was first done in~\cite{KO} and generalised in~\cite{cgkkm,Lee,obannon}. The formula obtained for the DC conductivity is given by
\begin{equation}\label{dbidc }
\sigma_{DC}=\sqrt{(\sigma_{DC}^{ccs})^2+(\sigma_{DC}^{diss})^2}\,.
\end{equation}
The first contribution, $\sigma_{DC}^{ccs}$, in the square root persists at zero charge density and is called charge conjugation symmetric (ccs) term following J.~Zaanen's suggestion. The second term is associated with the charge density and  is therefore due to the effect of dissipating momentum for charge carriers in the system. In the probe DBI case the momentum dissipation is due to the fact that charge degrees of freedom are subleading compared to uncharged ones. In such case, although there is a momentum conserving $\delta$-function, its coefficient is hierarchically suppressed. In the present paper we would like to generalise the work of Karch and O'Bannon~\cite{KO,obannon} that computes the DC conductivity in the presence of the constant finite electric and magnetic fields.

In this paper we will use the DBI action to investigate magneto-resistance in the presence of quantum critical behavior.
We find that within a suitable parameter space of dynamical critical exponent $z$ and hyperscaling violation exponent $\theta$ in the holographic theory, the in-plane magneto-resistivity can have the scaling behavior that is compatible  to that observed recently in experiments on BaFe$_2$(As$_{1-x}$P$_{x})_2$~\cite{analytis}.
Moreover, in such a case, the theory predicts a Hall resistivity in the same temperature regime that is linear in the magnetic field and approximately temperature independent.

As there are several possible holographic theories that reproduce this behavior, it would be interesting to have more data on other observables that could pin this theory down. AC resistivity data for example would be useful. A top-down realization of the appropriate exponents would be welcome.

The structure of this paper is as follows:

In the next section we introduce the holographic theory and derive its equations of motion. Section~\ref{sec:conduc} presents the conductivity and resistivity formulae obtained by the DBI theory. Section~\ref{sec:QC} addresses the black brane geometries that parametrize quantum criticality. In section~\ref{sec:scaling} we discuss the general scaling behavior of the magneto-resistivity with respect to temperature and magnetic field. We show the critical theories that reproduce the experimental observation~\eqref{linearTh} in section~\ref{sec:experiment} and consider some constraints in order to give a consistent parameter space of the holographic theory.
Section~\ref{sec:conclusion} contains our conclusion and outlook.

Several appendices are used to provide more technical details. We give the details of the computation of DC conductivity in appendix~\ref{DCconductivity} and discuss the DBI resistivity in various regimes in appendix~\ref{DBIres}. We show the hyperscaling-violating geometry can be supported by axions in appendix~\ref{HSV} and by a conserved charge density in appendix~\ref{HSVu1}. We discuss the validity of the probe approximation and the suppression of higher derivative corrections in appendix~\ref{probecondition} and appendix~\ref{highorder}, respectively. Appendix~\ref{app2scalars} devotes to general scaling solutions with two scalars and several massless vector fields.  In appendix~\ref{string} we focus on a top-down setup by using toroidal compactifications and discuss the corresponding DBI resistivity with respect to temperature and magnetic field.

\section{DBI Dynamics for Flavor Charge}

 We will consider the charged degrees of freedom of the holographic theory to be associated with the fundamental representations of the underlying fractionalized degrees of freedom. In string theory, string end-points transforming in the fundamental representation originate from open strings~\cite{book}. The conserved current will be dual to a U(1) gauge field in the bulk and its bulk dynamics will be controlled by the celebrated Dirac-Born-Infeld (DBI)  action and its variants.

The theory we will study involves a sector of charged carriers (the DBI sector) which interact amongst themselves as well as with a larger set of ``neutral" quantum critical degrees of freedom.\footnote{Here ``neutral" means the larger set of degrees of freedom are not charged under the flavor U(1).} That is to say, we will consider the theory which describes probe charge degrees of freedom interacting with a quantum critical bath in strongly coupled regimes. Furthermore, to connect to interesting experimental systems, we focus on 2+1 dimensional field theory, with a 3+1 dimensional bulk dual.

Based on above consideration, we introduce a bulk DBI action,
\begin{equation}
S_{DBI}=-\int d^{4}x \left[Z_1\sqrt{-\det(g_{\mu\nu}+Z_2F_{\mu\nu})}+{S\over 8}\e^{\m\n\r\s}F_{\m\n}F_{\r\s}\right]+\mathcal{O}(\partial F)\,,
\label{dbiaction}
\end{equation}
where the functions $Z_1,Z_2,S$ depend on (super)gravity  scalar fields originating in the closed string sector.
The CP-odd part of the action originates in the D-brane WZ terms~\cite{book}.
The subleading terms involve derivatives of the U(1) field strength. Intuition from the AdS/CFT correspondence indicates that they are suppressed in the strong coupling limit of the dual quantum field theory.

The dynamical field of this action is the U(1) gauge field that is dual to a conserved current in the quantum field theory. The action depends also on closed string fields: the space-time metric and other scalars. The DBI sector will be treated as a probe in a general metric background.\footnote{We will consider later the conditions that make such an approximation reliable.}
\begin{equation}
ds^2\equiv g_{\m\n}dx^{\mu}dx^{\nu}=-D(r)dt^2+B(r)dr^2+C(r) (dx^2+dy^2)\,.
\label{backgd}
\end{equation}
The scalars will be also functions of the radial direction only, making the coefficient functions $Z_1,Z_2,S$ depending on the holographic coordinate $r$.

We have included the CP-odd part as anomalies that are an important ingredient of the effective action for flavor. We  use the convention $\e^{rtxy}=+1$ for  the totally antisymmetric $\e^{\mu\nu\alpha\beta}$. The term $\mathcal{O}(\partial F)$ denotes the higher derivative corrections. We will consider the conditions for neglecting them, later on.

The ``charge neutral" quantum critical bath is partly holographically described by~\eqref{backgd}. We keep it general for the moment. There are also bulk scalars that appear in the coupling functions $Z_{1,2}$ and $S$. Later on we will finally consider the geometry which interpolates from an $AdS_4$ regime in the UV to a hyperscaling-violating regime in the IR, as motivated by condensed matter studies of quantum phase traditions in metals with weak Landau damping~\cite{Fitzpatrick:2013mja,Fitzpatrick:2013rfa,Hartnoll:2014gba,Allais:2014fqa}.

We are considering the system that is at finite charge density as well as a constant magnetic field $h$.
We will calculate the conductivity upon turning on a finite electric field in the $x$ direction and a magnetic field $h$ perpendicular to the $x$-$y$ plane.
This calculation was done first in \cite{KO,obannon} and we will generalize it here.\footnote{These results were first obtained in 2011 with Bom Soo Kim but were not published.}

We take the following ansatz for world-volume gauge fields
\begin{equation}\label{dbiansatz}
A_t=a_t(r)\sp A_x=-E\,t+a_x(r)\sp A_y=h\,x+a_y(r)\,,
\end{equation}
with $E$ the electric field and $h$ the magnetic field. The action~\eqref{dbiaction} now only depends on $r$-derivatives of gauge fields.
We therefore have three constants of motion that can be interpreted as the expectation values of the (dual) current components:
\be
{Z_1Z_2^2[C^2a_t'-Z_2^2h(E a_y'-ha_t')]\over \sqrt{X}}+Sh=\t\,,
\label{eomjt}
\ee
\be
-{Z_1Z_2^2\,C\,Da_x'\over\sqrt{X}}=\x\,,
\label{x}
\ee
\be
-{Z_1Z_2^2\,\left[DCa_y'-Z_2^2E(Ea_y'-ha_t')\right]\over\sqrt{X}}-SE=\y\,,
\label{eomjxjy}
\ee
with
\be
X=DBC^2-Z_2^2\left[C^2a_t'^2+BCE^2-DC(a_x'^2+a_y'^2)-DBh^2\right]-Z_2^4(Ea_y'-ha_t')^2
\label{eomX}\,.
\ee
Here primes denote derivatives with respect to $r$. As mentioned above, we identify the three constants $\t,\y,\x$ as the dual current $\langle J^\mu\rangle=\frac{\delta S_{DBI}}{\delta A_\mu}$, where $\mu=t, x, y$. In particular, $\t$ is the charge density in dual field theory, while $\y,\x$ are the currents induced by the electric field in the presence of the magnetic field.

\section{DBI Conductivity}\label{sec:conduc}

We now proceed by solving the equations~\eqref{eomjt}-\eqref{eomjxjy} for the unknown functions $a_t(r),a_x(r),a_y(r)$ in terms of three current expectation values and obtain the on-shell action. In holography, a regularity condition typically gives the relation between the sources (here the charge density $\t$, and the electric and magnetic fields ($E,h$)) and the expectation values $\x,\y$. This regularity condition can be obtained by demanding the action to be real along the lines of~\cite{KO,obannon}. The detailed calculation is carried out in appendix~\ref{DCconductivity}, and the resulted conductivity (in the limit of a vanishing electric field) is given by
\be
\s_{xy}={h~Z_2(r_0)^2\t+C(r_0)^2S(r_0)\over C(r_0)^2+Z_2(r_0)^2h^2}\,,
\label{sigxy}
\ee
\be
\s_{xx}={Z_2(r_0)C(r_0)\sqrt{Z_1(r_0)^2Z_2(r_0)^2(C(r_0)^2+Z_2(r_0)^2h^2)+(\t-S(r_0)h)^2}\over C(r_0)^2+Z_2(r_0)^2h^2}\,,
\label{sigxx}
\ee
with $\t$ the charge density and $r_0$ the location of the bulk black-hole horizon.\footnote{If the electric field is finite, then $r_0$ is replaced by another point $r_s$ that is determined by \eqref{21} in appendix~\ref{DCconductivity}.}
The four functions that enter the conductivity are the couplings $Z_{1,2}, S$ and the metric component $C$ of the bulk metric. All four are functions of the holographic radial coordinate and they are evaluated at the horizon.
As $r_0$ is in general a function of the temperature $T$, the conductivity is a function of $(T, h)$ as well as the charge density $\t$.

Note also that the Hall conductivity~\eqref{sigxy} is proportional to the CP-breaking terms $(h, S(r_0))$.
We may also compute the resistivity matrix by inverting the conductivity matrix,
\be\label{rex}
\rho_{xx}=\frac{\sigma_{xx}}{\sigma_{xx}^2+\sigma_{xy}^2}=\frac{CZ_2\sqrt{{(\t-h S)^2}+C^2 Z_1^2 Z_2^2+h^2 Z_1^2 Z_2^4}}{\t^2 Z_2^2+C^2Z_1^2 Z_2^4+C^2 S^2}\,,
\ee
\be
\rho_{xy}=-\frac{\sigma_{xy}}{\sigma_{xx}^2+\sigma_{xy}^2}=-\frac{h\,\t Z_2^2+C^2S}{\t^2Z_2^2+C^2 Z_1^2 Z_2^4+C^2 S^2}\,,
\label{rey}
\ee
with all functions evaluated at the horizon $r_0$. From now on we will not explicitly indicate the $r_0$ dependence.

In appendix \ref{DBIres} we investigate various limits of the resistivity and conductivity formulae \eqref{sigxy}-\eqref{rey}.  Here we will consider only the case  of time (T)-invariant theories where $S=0$. The conductivity in this case at zero magnetic field reads
\be
\s_{xx}={Z_2\over C}\sqrt{Z_1^2Z_2^2~C^2+\t^2}\sp \sigma_{xy}=0\,.
\label{ss1}
\ee
There are two relevant regimes:
\begin{itemize}
\item The Drude regime (DR) when $Z_1Z_2~C\ll |\t|$ with the conductivity given by
\be
\s_{xx}\simeq {Z_2\over C}|\t|\,.
\label{ss2}  \ee

\item The Charge-conjugate regime (CCR) when $Z_1Z_2~C\gg |\t|$ with     conductivity given by
\be
  \s_{xx}\simeq Z_1Z_2^2\,.
\label{ss3}
\ee
\end{itemize}
Reinstating the magnetic field,  the formulae for the resistivity in the two different regimes become:
\begin{itemize}
\item In the  Drude regime (DR)
\be
  \s_{xx}\simeq {Z_2\over C}|\t|{\sqrt{1+{Z_1^2Z_2^4\over \t^2}h^2}\over 1+{Z_2^2\over C^2}h^2}\sp \s_{xy}={Z_2^2\t\over C^2+Z_2^2h^2}h\,,
\label{s4}
\ee
\be
\rho_{xx}\simeq {{C\over \t^2 Z_2}}\sqrt{\t^2+Z_1^2Z_2^4~h^2}\sp \rho_{xy}\simeq {-{h\over \t}}\,,
\label{s6}\ee
while the inverse Hall angle becomes
\be
\cot \Theta_H\equiv {\s_{xx}\over \s_{xy}}\simeq {C\over Z_2~h}\sqrt{1+{Z_1^2Z_2^4\over \t^2}h^2}={C\over Z_2~h}+{\cal O}(h)\,.
\label{s8}
\ee
\item The Charge-conjugate regime (CCR) when $Z_1Z_2\,C\gg |\t|$ with conductivity given by
\be
 \s_{xx}\simeq Z_1Z_2^2{C\over \sqrt{C^2+Z_2^2~h^2}}\sp \s_{xy}={Z_2^2\t\over C^2+Z_2^2h^2}h\,,
\label{s5}
 \ee
\be
\rho_{xx}\simeq {\sqrt{C^2+Z_2^2~h^2}\over Z_1Z_2^2C}\sp \rho_{xy}\simeq {-{\t ~h\over Z_1^2Z_2^2C^2}}\,,
\label{S7}\ee
\be
\cot\Theta_H\simeq {Z_1C\over \t~h}\sqrt{C^2+Z_2^2h^2}={Z_1C^2\over \t~h}+{\cal O}(h)\,.
\label{s9}
\ee
\end{itemize}

\section{Parametrizing Quantum Criticality \label{sec:QC}}

We will now restrict our attention to theories for which the closed string sector is quantum critical in a generalized sense by allowing hyperscaling violation. Assuming translational and rotational invariance in space and time, the criticality can be holographically described by a metric that is hyperscaling violating at zero temperature~\cite{cgkkm,qc}.
\begin{equation}\label{fixpoint}
ds^2_{\text{IR}}=r^\theta\left(-\frac{dt^2}{r^{2z}}+\frac{L^2 dr^2}{r^2} +\frac{dx^2+dy^2}{r^2}\right).
\end{equation}
 This geometry is characterised by two parameters: the dynamical critical exponent $z$ and the hyperscaling violation exponent $\theta$. The geometry is  generically singular but many such metrics satisfy Gubser's physicality criterion,\footnote{A detailed discussion of the meaning of such a criterion as well as related ones can be found in \cite{cgkkm}.} which suggests that the singularity is resolvable and   that restricts the parameter space of $(z,\theta)$. The allowed parameter range is given by
\begin{eqnarray}
\text{IR}\quad r\rightarrow 0&:& [z\leqslant 0, \theta>2],\quad [0<z<1, \theta>z+2]\,,\label{irlocates0}\\
\text{IR}\quad r\rightarrow \infty&:& [1<z\leqslant 2, \theta<2z-2],\quad [z>2, \theta<2]\label{irlocates}\,.
\end{eqnarray}
What this means is that in the case~\eqref{irlocates0}, the IR region of the metric is near $r\to 0$, while in the case~\eqref{irlocates} the IR region of the metric is near $r\to \infty$.

At  finite temperature, the associated black hole metric can be written as
\begin{equation}\label{fixpointT}
ds^2_{\text{IR}}=r^\theta\left(-f(r)\frac{dt^2}{r^{2z}}+\frac{L^2 dr^2}{f(r)}+\frac{dx^2+dy^2}{r^2}\right),
\end{equation}
where the precise form of the blackness function $f(r)$ depends on the theory one is considering. At the horizon $r_0$, $f(r_0)=0$ and  we obtain $r_0\sim T^{-{1\over z}}$.

Such black hole geometries can be supported by a conserved charge density (in our case here this should be different than the DBI charge density whose conductivity we are studying), see appendix~\ref{HSVu1} for more detail. They can also be supported by scalars without a potential (axions) and such solutions are presented in appendix~\ref{HSV}.

The validity of the probe flavor description demands that the stress tensor of DBI action is much smaller than the one that is used to seed the background geometry~\eqref{fixpointT}. We will discuss the regime of validity in section~\ref{sec:experiment}.


\section{Quantum Criticality and Magneto-resistance}\label{sec:scaling}

We will now assume that the bulk metric is given by the scaling form (\ref{fixpointT}) and that a bulk scalar $\phi$ is running to support this solution as described in appendix \ref{HSV} and in appendix \ref{HSVu1}.

We will also parametrize the scalar functions in the DBI action (to leading order in the IR expansion) in terms of the running scalar as
\be
Z_1=Z_{10}~e^{a\,\phi}\sp Z_2=Z_{20}~e^{b\,\phi}\sp S=S_0~e^{c\,\phi}\,.
\ee
This parametrization is in accordance with string theory calculations in a generic D-brane setting~\cite{book}.
From~\eqref{rex}, the magneto-resistance reads
\begin{equation}
\rho_{xx}=\frac{C(r_0)Z_2(r_0)\sqrt{{(\t-h S(r_0))^2}+C(r_0)^2 Z_1(r_0)^2 Z_2(r_0)^2+h^2 Z_1(r_0)^2 Z_2(r_0)^4}}{Z_2(r_0)^2\left[\t^2 +C(r_0)^2Z_1(r_0)^2 Z_2(r_0)^2\right]+C(r_0)^2S(r_0)^2}\;.
\label{22}
\end{equation}
In the scaling solutions we have
\be
C\sim r_0^{\theta-2}\sp e^{\phi}\sim r_0^{\kappa_{\pm}}\sp  r_0\sim T^{-{1\over z}}\,.
\ee
The value of $\kappa_{\pm}$ depends on the type of bulk solution.
Whether the bulk background is supported by massless scalars as in appendix \ref{HSV}   or by a charge density as in appendix \ref{HSVu1}, we obtain
\be
\kappa_{\pm}=\pm \sqrt{(\theta-2)(\theta+2-2z)}\,.
\ee
After substituting into~\eqref{22}, we obtain the scaling form
\be
\rho_{xx}\sim T^{\lambda_1}{\sqrt{{(\t-c_0\, h \,T^{\lambda_0})^2}+c_1 ~T^{\lambda_2}+c_2 ~h^2~ T^{\lambda_3}}\over \t^2+c_1 ~T^{\lambda_2}+c_3~T^{\l_T}}\,,
\label{cc1}
\ee
where $c_{0,1,2,3}$ are constants and the exponents are given by
\begin{equation}\label{relation}
\begin{split}
{ \lambda_0=-\frac{c\,\kappa_{\pm}}{z}},\quad \lambda_1={b\kappa_{\pm}+2-\theta\over z}
\sp
{\lambda_2={2(2-\theta)-2(a+b)\kappa_{\pm}\over z}}\,,\\
{\lambda_3=-{2(a+2b)\kappa_{\pm}\over z}}\sp
{\lambda_T={2(2-\theta)-2(c-b)\kappa_{\pm}\over z}}\,.
\end{split}
\end{equation}

The Hall conductivity
\be
\rho_{xy}=-\frac{h\,\t Z_2^2+C^2S}{\t^2Z_2^2+C^2 Z_1^2 Z_2^4+C^2 S^2}\,,
\ee
becomes
\be
\rho_{xy}\sim{h\,\t+c_4~T^{\l_4}\over \t^2+c_1 ~T^{\lambda_2}+c_3~T^{\l_T}}\,,
\label{cc2}
\ee
with
\be
{\l_{4}={2(2-\theta)-(c-2b)\kappa_{\pm}\over z}}\,.
\ee

When the intrinsic T-violation vanishes then {$c_0=c_3=c_4=0$}.
From now on we will consider the theory to be T-invariant and therefore we set  {$c_0=c_3=c_4=0$}.

The critical formulae for the conductivities (\ref{cc1}) and (\ref{cc2}) simplify further in the two distinct regimes we defined earlier.

\begin{itemize}
\item In the Drude regime they become
\be
\rho_{xx}\sim {{T^{\lambda_1+{\l_3\over 2}}}\over \t^2}\sqrt{\t^2~T^{-\l_3}+c_2 ~h^2} \sp \rho_{xy}\sim{h\over \t}\,.
\label{cc11}
\ee

\item In the Charge-conjugate regime they are instead

\be
\rho_{xx}\sim T^{{\lambda_1-\lambda_2+{\l_3\over 2}}}\sqrt{c_1 ~T^{\lambda_2-\l_3}+c_2 ~h^2}\sp \rho_{xy}\sim {h\t\over T^{\lambda_2}}\,.
\label{cc3}
\ee

\end{itemize}

The scaling exponents appearing in the formulae for magneto-resistance~\eqref{cc1} and~\eqref{cc2}, namely $z,\theta$ as well as $\l_{1,2,3,4,T}$ are continuous adjustable parameters at the level of effective holographic theory. They are however discrete parameters in string theory, and they depend on the type of string ground state (compactification) considered as well as the type of brane and its embeddings.

\section{Revisiting the Critical Magneto-resistance of Pnictides}\label{sec:experiment}

The measurements on BaFe$_2$(As$_{1-x}$P$_{x})_2$ near the optimal doping imply that the in-plane magneto-resistance can be described to good accuracy by~\eqref{linearTh}.
In this section, we will find critical theories  that can reproduce the recent experimental observation in~\cite{analytis}. We set $S=0$ as the material in question does not seem to have intrinsic CP-violation.

There are three possible ways to realize this scaling form~\eqref{linearTh} in our setup:

\begin{enumerate}

\item In the Drude regime by having
\be
 \lambda_1=1\sp  \lambda_3=-2\,.
\ee
The regime of validity of the scaling form is as $T\to 0$ when $\lambda_2>0$ or as $T\to \infty$ when $\l_2<0$.
This will be realized if
\be
{a=\frac{2(2-\theta)-z}{\kappa_{\pm}}\sp b=\frac{\theta+z-2}{\kappa_{\pm}}}\,,
\ee
which further determines $\lambda_2=0$. In this case the Hall resistivity is temperature independent.
\be
{\rho_{xy}=-{h\over \t}}\,.
\ee

\item  In the Charge-conjugate regime if
\be
 {\lambda_1={1\over 2}\lambda_3+2}\sp \lambda_2=\lambda_3+2\,,
\ee
from which we obtain
\be
{ a=\frac{2(2-\theta)-z}{\kappa_{\pm}},\quad b=\frac{\theta+z-2}{\kappa_{\pm}}}\,.
\ee
Using above result as well as~\eqref{relation}, we then find
\be
{\l_1=1\sp \l_2 =0\sp \l_3=-2}\,.
\ee

\item In any regime if
\be
{\l_1=1\sp \l_2 =0\sp \l_3=-2}\,,
\ee
or
\be
{ a=\frac{2(2-\theta)-z}{\kappa_{\pm}},\quad b=\frac{\theta+z-2}{\kappa_{\pm}}}\,.
\label{condi}
\ee

Interestingly, we note that if the scaling appears in any of the two regimes  then  it is valid in all regimes.

The conditions~\eqref{condi} imply in terms of the background data that
\begin{equation}\label{linearcond}
\frac{C(r_0)}{Z_2(r_0)}=\frac{T}{\ell_0}\sp Z_1(r_0)Z_2(r_0)^2=\frac{Z_0}{T}\,,
\end{equation}
where $\ell_0, Z_0$ are two positive constants. Then $\rho_{xx}$ becomes
\begin{equation}\label{hololinearT}
\rho_{xx}=\frac{Z_0}{\ell_0^2\t^2+Z_0^2}\sqrt{\left(\frac{\ell_0^2\t^2}{Z_0^2}+1\right)T^2+\ell_0^2\,h^2}\,.
\end{equation}
By  introducing dimensionless temperature and magnetic field variables $\textbf{T}, \textbf{h}$ as well as a constant $\tau$
\begin{equation}\label{dimenless}
\textbf{T}=\frac{T}{Z_0}\sp \textbf{h}=\frac{h}{\t}\sp \tau=\frac{Z_0}{\ell_0 \t}\,,
\end{equation}
 we may rewrite the DBI resistivity $\rho_{xx}$ as
\begin{equation}\label{hololinearTh}
\rho_{xx}=\frac{|\tau|}{1+\tau^2}\sqrt{(1+\tau^2)\textbf{T}^2+\textbf{h}^2}\;.
\end{equation}
Moreover, the Hall resistivity can be obtained from~\eqref{rey} by setting $S=0$ and substituting from~\eqref{linearcond}.
\begin{equation}
\rho_{xy}=-\frac{\sigma_{xy}}{\sigma_{xx}^2+\sigma_{xy}^2}=-\frac{\textbf{h}}{1+\tau^2}\,.
\end{equation}
It is linear in the magnetic field and is temperature independent.

To explore these observations in the rest of this section and with the assistance of several appendices, we will address the following question:
For theories satisfying~\eqref{linearcond}, when are the approximations we made to derive the DBI resistivity valid? There are two approximations that are relevant to this question.

(a) The first concerns the back-reaction of the DBI solution to the bulk equations of motion (that has been neglected).

(b) The second concerns whether the higher derivative corrections to the DBI action can be neglected.

\end{enumerate}

For the first question,  we are assuming that the bulk geometry is of the scaling type described in~\eqref{fixpointT}. As mentioned above, such geometries can be reliable generated by bulk holographic theories with bulk massless scalars (see appendix~\ref{HSV}) or by a vector potential (see~\cite{cgkkm} and appendix~\ref{HSVu1}). The question in this case boils down to whether the values of the critical exponents ($z,\theta$) needed, satisfy the physicality constraints, notably the Gubser bound.

As the conditions~\eqref{linearcond} also impose constraints on the DBI coupling functions $Z_1(\phi)$, $Z_2(\phi)$, the related relevant question is whether the required functions can be obtained from known constructions in string theory. This issue is partly addressed  in appendix~\ref{string} where simple toroidal compactifications are analysed.

What we need to do next is to check whether we can find physically reasonable background exponents $(z,\theta)$ that allow (\ref{linearcond}). For concreteness we focus on the hyperscaling-violating geometries~\eqref{hyperscal}. According to our previous discussion there are three kinds of constraints we should consider.

\begin{itemize}
  \item 1) To resolve the naked singularity (Gubser criterion) which is present in the hyperscaling-violating ground state~\eqref{fixpoint}, we obtain the parameter range~\eqref{irlocates0} and~\eqref{irlocates}.

  \item 2) We couple the DBI action to the bulk theory, but treat it as a probe. The validity of the probe approximation imposes the constraint~\eqref{probe}. As discussed in appendix~\ref{probecondition}, we obtain two consistent cases:
 \begin{eqnarray}
\text{low temperature limit} &:&\textbf{T}\ll 1\quad \text{with}\quad \frac{\theta+z-4}{z}>0\,, \\
\text{high temperature limit}&:& \textbf{T}\gg 1\quad \text{with}\quad \frac{\theta+z-4}{z}<0\,.
\end{eqnarray}

Notice that the constraint involved the dimensionless temperature in (\ref{dimenless}). It was rendered dimensionless using the scale present due to hyperscaling violation. This scale should be thought of as the scale of an irrelevant coupling in the IR. Therefore, even in the case $\textbf{T}\gg 1$ could be in reality a ``low" temperature regime provided the hyperscaling-violating scale is lower than the temperatures of the experiments.

  \item 3) The higher derivative corrections $\mathcal{O}(\partial F)$ in the action~\eqref{dbiaction} should be sufficiently small compared with the leading order term $F_{\mu\nu}F^{\mu\nu}$. We need to consider the low temperature case $\textbf{T}\ll 1$ and the high temperature case $\textbf{T}\gg 1$, and the allowed parameter space is given by
 \begin{eqnarray}
\textbf{T}\ll 1&:&\quad \theta/z>0\,, \\
\textbf{T}\gg1&:&\quad \theta/z<0\,.
\end{eqnarray}
One can also choose $\theta=0$ in very special case. For more details one can consult for appendix~\ref{highorder}.
\end{itemize}

\begin{figure}[ht!]
\begin{center}
\includegraphics[width=.45\textwidth]{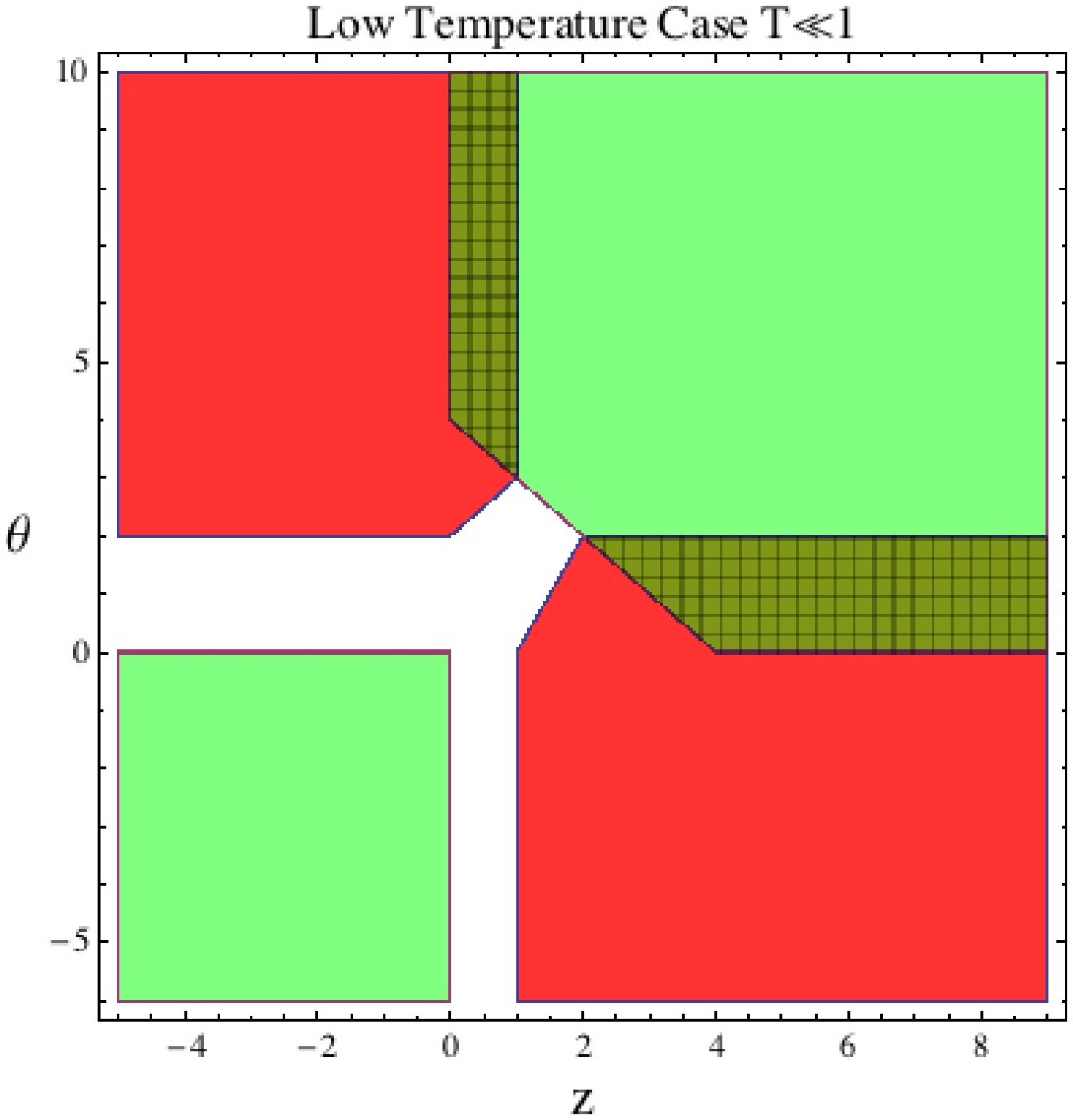}\quad\quad
\includegraphics[width=.45\textwidth]{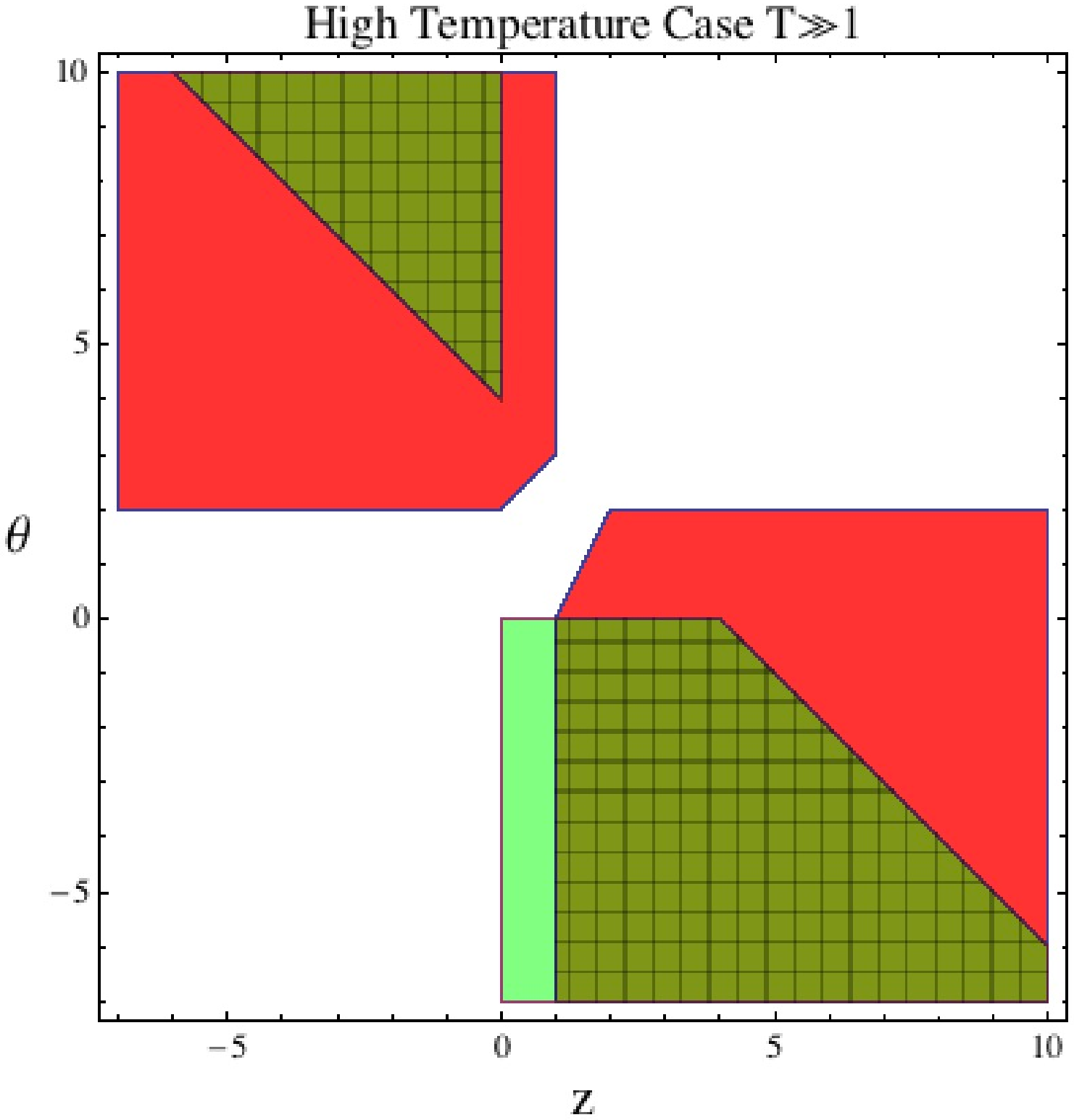}
\caption{Parameter space for quantum criticality. (a) left $\textbf{T}\ll1$, (b) right $\textbf{T}\gg 1$. The red region is the parameter space constrained by Gubser criterion. The green part is the region in which the probe approximation and the suppression of higher derivative corrections are valid under the condition~ (\protect\ref{linearcond}). Therefore, the overlap regime with grids gives the parameter range that satisfies all constraints.}
\label{fig:space}
\end{center}
\end{figure}

Those constraints are important in order to give a consistent parameter space of the holographic theory. By considering all three constraints, the allowed parameter range is given by
\begin{eqnarray}\label{linearTH}
\label{lineara}\textbf{T}\ll1 &:& [0<z< 1, \theta>4-z]\; (\text{IR}\;\;r\rightarrow 0),\\
 &&[2<z\leqslant 4, 4-z<\theta<2] \;[z>4,0<\theta<2]\;(\text{IR}\;\;r\rightarrow \infty)\\
\label{linearTH}\textbf{T}\gg1&:& [z< 0, \theta>4-z]\; (\text{IR}\;\;r\rightarrow 0),\\
 \label{lineard}&&[1<z\leqslant 4,\theta<0] \;[z>4,\theta<4-z]\;(\text{IR}\;\;r\rightarrow \infty).
\end{eqnarray}
The corresponding parameter range is shown in figure~\ref{fig:space}. The parameter $(\theta, z)$ in the regime with grids satisfies all  constraints we mentioned above.
By further choosing suitable couplings $Z_1$ and $Z_2$ that satisfy~\eqref{linearcond}, we can obtain the expected magnetoresistance~\eqref{hololinearTh} from the DBI theory.

\section{Conclusion and Outlook}\label{sec:conclusion}
A simple holographic system, namely flavor charge carriers described by a D-brane DBI action coupled with hyperscaling-violating geometries, provides a solvable quantum critical model of magneto-transport. An interesting property emerging from our work is the scaling behavior of the magneto-resistivity with respect to temperature and magnetic field.

Within a suitable parameter space of dynamical critical exponent $z$ and hyperscaling violation exponent $\theta$ in the holographic theory, the in-plane magneto-resistivity has been shown to have scaling behavior that is compatible  to that observed recently in experiments on BaFe$_2$(As$_{1-x}$P$_{x})_2$.
Moreover, in such a case, the theory predicts a Hall resistivity in the same temperature regime that is linear in the magnetic field and approximately temperature independent.

As one can see from figure~\ref{fig:space}, there are several possible holographic theories that reproduce the behavior~\eqref{linearTh}, it would be interesting to have more data on other observables that could pin this theory down. AC resistivity data for example would be useful.
The down side of the effective holographic theories proposed here is that there is no clear indication of what the microscopic description is for the strongly coupled quantum critical theory in terms of material-related variables. A top-down realization of the appropriate exponents would be helpful.

An extension to current work would be to understand the underlying dynamics and to match the holographic description with the expected interactions of electrons in real materials. A first step towards this goal has been addressed~\cite{sachdev} and connected to phases and critical points of Hubbard model~\cite{Sachdev:2010uz}.

\addcontentsline{toc}{section}{Acknowledgments}
\acknowledgments\label{ACKNOWL}

We would like to thank James Analytis for explaining the results of \cite{analytis} to us and Jan Zaanen for discussions. E. K. would like to thank Bom Soo Kim as some of the results presented here were derived with him in earlier unpublished work.

This work was supported in part by European Union's Seventh Framework Programme under grant agreements (FP7-REGPOT-2012-2013-1) no 316165 and the Advanced ERC grant SM-grav 669288.

\appendix
\renewcommand{\theequation}{\thesection.\arabic{equation}}
\addcontentsline{toc}{section}{Appendix}
\section*{Appendices}


\section{Detailed Calculation of DC Conductivity}\label{DCconductivity}
In this appendix, we calculate the conductivity tensor by using techniques developed in~\cite{KO,obannon}. The calculation is done in the probe approximation on top of the background~\eqref{backgd}. Before computing the conductivity, we first fix the dimensions of various quantities.

From the supergravity point of view, the original DBI action can be given by
\begin{equation}
S_{DBI}=-M_p^4\int d^{4}x \left[\mathcal{Z}_1\sqrt{-\det(g+\mathcal{Z}_2\ell_s^2 F)}+{\ell_s^4 \mathcal{S}\over 8}\e^{\m\n\r\s}F_{\m\n}F_{\r\s}\right],
\end{equation}
with $\ell_s$  the string length and $M_p$ the Planck mass. The dimension of each quantity is given as follows:
\begin{equation}
[\mathcal{Z}_1]=[\mathcal{Z}_2]=[\mathcal{S}]=0\sp [M_p]=[A_\mu]=[1/\ell_s]\,.
\end{equation}
To simplify the action, we introduce new coupling functions
\begin{equation}
Z_1=M_p^4 \mathcal{Z}_1\sp Z_2=\ell_s^2 \mathcal{Z}_2\sp S=M_p^4 \ell_s^4\mathcal{S}\,,
\end{equation}
We obtain the action~\eqref{dbiaction} with dimensions
\begin{equation}
[Z_1]=[1/\ell_s^4]\sp [Z_2]=[\ell_s^2]\sp [S]=0\,.
\end{equation}

Under the ansatz~\eqref{dbiansatz} the field strengths are only $r$-dependent, and we obtain the three conserved charges $(\t, \x, \y)$ as shown in~\eqref{eomjt}-\eqref{eomjxjy}.  They are the currents in the dual field theory. Here we have assumed that the dual field theory is located at $r\rightarrow\infty$.

To move forward, we further define the following combinations
\be
\xi=DC^2 + Z_2^2(D\,h^2 - CE^2)\sp \a=Z_2^2\,[hD\tt +CE\yy]\,,
\label{eomxialpha}
\ee
\be
\chi=DC^{2}Z_1^2Z_2^4 +Z_2^2\,[D\tt^2 - C(\x^2 + \yy^2)]\,.\label{eomchi}
\ee
The equations of motion~\eqref{eomjt}-\eqref{eomjxjy} are then reduced to
\be
a_t'={\sqrt{BD}\over C}~{\tt \xi - h\a\over \sqrt{\xi\chi - \a^2}}~ {|\xi|\over\xi}\,,
\label{dbisoluat}\ee
\be
a_x'=-\sqrt{B\over D}~{\x|\xi|\over \sqrt{\xi\chi - \a^2}}\,,
\label{dbisoluax}\ee
\be
a_y'=-\sqrt{B\over D}~{\yy\xi+E\a\over \sqrt{\xi\chi - \a^2}}~{|\xi|\over\xi}\,,
\label{dbisoluay}
\ee
with
\be
\sqrt{X}=Z_1Z_2^2C\sqrt{DB}{|\xi|\over \sqrt{\xi\chi-\a^2}}\,.
\label{dbisolux}
\ee
We can obtain the on-shell action as
\be\label{onshell}
S_{DBI}^{\rm on-shell}=-\int d^4x ~\left[Z_1^2Z_2^2C\sqrt{DB} \sqrt{\xi^2\over \xi\chi-\a^2 }\right.
\ee
$$
\left.-{S\sqrt{BD}\over  CD\sqrt{\xi\chi - \a^2}}{|\xi|\over\xi}\left[(D\t h+C\y E)\xi+(CE^2-Dh^2)(S\xi+\a)\right]\right].
$$

Note that what we are considering is a rotationally-invariant system with electric field $E$ in the $x$ direction and magnetic field $h$ perpendicular to the $x$-$y$ plane. Therefore we have $\s_{xx}=\s_{yy}$ and $\s_{xy}=-\s_{yx}$. $\s_{xx}$ is the Ohmic conductivity and the off-diagonal component $\s_{xy}$ is Hall conductivity.

To determine the current responses to the external fields, we focus on the square root appearing in the on-shell action~\eqref{onshell}. As in~\cite{KO,obannon}, we find that demanding reality of the on-shell action allows us to solve for $\x$ and $\y$ in terms of $E, h$ and $\t$, and hence the conductivity.

The detailed argument is as follows.
We observe  from~\eqref{eomxialpha} that $\xi$ must vanish somewhere between the horizon where $D=0$ and the AdS boundary.
 The reason is that it is positive at the boundary, since $D,C\to\infty$ while $E,h,Z_2$ are or become constants and the first term dominates.
 The same function is  negative at the horizon as there $D\to 0$ while all other quantities asymptote to constants.
 We will denote by $r_s$ the zero of the function  $\xi$ and by $r_0$ the position of the horizon.\footnote{As the argument implies, multiple zeros do not give further solutions.} The same applies to $\chi$. It is positive at the AdS boundary as the first term dominates and it is negative at the horizon as all terms multiplied by $D$ vanish.

We will now argue that the zero of the function $\chi$,  say $r_t$,  must coincide with the zero of $\xi$, $r_s$, in order for the square root to be real.
Indeed assume that $r_t\not= r_s$, then in the interval $r_t<r<r_s$, $\xi\chi<0$ and the square root $\sqrt{\xi^2\over \xi\chi-\a^2 }$ is imaginary. The same applies if $r_s<r_t$. We therefore conclude that $r_s=r_t$.

We can now argue that $\a$ must also vanish at $r_s$. The reason is that if $\a(r_s)\not=0$ then at $r_s$ the square root in the second part of the action  $\sqrt{\xi\chi -\a^2 }\to \sqrt{-\a(r_s)^2}$ becomes imaginary. Therefore regularity implies that $\a(r_s)=0$. In conclusion,
\be
\xi(r_s)=\chi(r_s)=\alpha(r_s)=0\,.
\ee
The first equation
\be
\xi(r_s)=D(r_s)C(r_s)^2 + Z_2^2(r_s)(D(r_s)h^2 - C(r_s)E^2)=0\,,
\label{21}
\ee
determines the vanishing point $r_s$.
Solving the other two equations, namely $\chi(r_s)=\alpha(r_s)=0$, we obtain
\be
\y=-{h~Z_2(r_s)^2\t+C(r_s)^2S(r_s)\over C(r_s)^2+Z_2(r_s)^2h^2}~E\,,
\label{jxex}
\ee
\be
\x={Z_2(r_s)C(r_s)\sqrt{Z_1(r_s)^2Z_2(r_s)^2(C(r_s)^2+Z_2(r_s)^2h^2)+(\t-S(r_s)h)^2}\over C(r_s)^2+Z_2(r_s)^2h^2}~E\,.
\label{jyex}
\ee
We can read off the DC conductivity as
\be
\s_{xy}={h~Z_2(r_s)^2\t+C(r_s)^2S(r_s)\over C(r_s)^2+Z_2(r_s)^2h^2}\,,
\label{sigxya}
\ee
\be
\s_{xx}={Z_2(r_s)C(r_s)\sqrt{Z_1(r_s)^2Z_2(r_s)^2(C(r_s)^2+Z_2(r_s)^2h^2)+(\t-S(r_s)h)^2}\over C(r_s)^2+Z_2(r_s)^2h^2}\,.
\label{sigxxa}
\ee
These are the fully non-linear conductivities.

To evaluate the conventional conductivity at vanishing electric field, $E\to 0$, we must return to equation (\ref{21}) and set $E\to 0$ to obtain
\be
D(r_s)\left[C(r_s)^2 + Z_2^2(r_s)h^2\right]=0\,,
 \ee
 which implies that $D(r_s)=0$. The metric function $D(r_s)$ is non-negative and vanishes only at the horizon so that
\be
E\to 0\sp r_s\to r_0\,.
\ee
Therefore, to obtain the conventional conductivity in the absence of an electric field we must evaluate all functions in~\eqref{sigxya} and~\eqref{sigxxa} on the horizon.\footnote{Note that here we assumed that the UV is at $r\rightarrow\infty$. If the IR is at $r\rightarrow\infty$ instead, we make the replacement $(\t,\x,\y,S)\rightarrow-(\t,\x,\y,S)$, then the main results we are interested, especially $\sigma_{xx}$, will not change. The only difference is that $\sigma_{xy}\rightarrow-\sigma_{xy}$.}

\section{DBI Resistivity in Various Regimes}\label{DBIres}

In this appendix we will analyse the DC resistivity as derived in the previous appendix, in various regimes.

We start from the DC conductivity as derived in the previous appendix
\be
\s_{xy}={h~Z_2(r_0)^2\t+C(r_0)^2S(r_0)\over C(r_0)^2+Z_2(r_0)^2h^2}\,,
\label{sigxy1}
\ee
\be
\s_{xx}={Z_2(r_0)C(r_0)\sqrt{Z_1(r_0)^2Z_2(r_0)^2(C(r_0)^2+Z_2(r_0)^2h^2)+(\t-S(r_0)h)^2}\over C(r_0)^2+Z_2(r_0)^2h^2}\,,
\label{sigxx1}
\ee
with $\t$ the charge density and $r_0$ the location of the bulk black brane horizon.

Note that the Hall conductivity~\eqref{sigxy1} is proportional to the CP-breaking terms $(h, S(r_0))$.
We may also compute the resistivity matrix by inverting the conductivity matrix,
\begin{eqnarray}\label{rex1}
\rho_{xx}&=&\frac{\sigma_{xx}}{\sigma_{xx}^2+\sigma_{xy}^2}=\frac{CZ_2\sqrt{(\t-h\,S)^2+C^2 Z_1^2 Z_2^2+h^2 Z_1^2 Z_2^4}}{\t^2 Z_2^2+C^2Z_1^2 Z_2^4+C^2 S^2}\,,\\
\rho_{xy}&=&-\frac{\sigma_{xy}}{\sigma_{xx}^2+\sigma_{xy}^2}=-\frac{h\,\t Z_2^2+C^2S}{\t^2Z_2^2+C^2 Z_1^2 Z_2^4+C^2 S^2}\,,
\label{rey1}
\end{eqnarray}
with all functions evaluated at the horizon $r_0$. From now on we will  not explicitly indicate the $r_0$ dependence.

We will investigate  various limits of the general conductivity formula, and we will start by considering first the case of T-invariant theories where $S=0$. The conductivity in this case at zero magnetic field reads:
\be
\s_{xx}={Z_2\over C}\sqrt{Z_1^2Z_2^2~C^2+\t^2}\sp \sigma_{xy}=0\,.
\label{sss1}
\ee
There are two possibilities:
\begin{itemize}
\item The Drude regime (DR) when $Z_1Z_2~C\ll |\t|$ with the conductivity given by
\be
  \s_{xx}\simeq {Z_2\over C}|\t|\,.
\label{sss2}  \ee

\item The Charge-conjugate regime (CCR) when $Z_1Z_2~C\gg |\t|$ with     conductivity given by
\be
  \s_{xx}\simeq Z_1Z_2^2\,.
\label{sss3}  \ee

\end{itemize}

After we reinstate the magnetic field, the formulae for the resistivities in the two different regimes become:
\begin{itemize}
\item In the  Drude regime (DR)
\be
  \s_{xx}\simeq {Z_2\over C}|\t|{\sqrt{1+{Z_1^2Z_2^4\over \t^2}h^2}\over 1+{Z_2^2\over C^2}h^2}\sp \s_{xy}={Z_2^2\t\over C^2+Z_2^2h^2}h\,,
\label{ss4}  \ee
\be
\rho_{xx}\simeq {C\over \t^2 Z_2}\sqrt{\t^2+Z_1^2Z_2^4~h^2}\sp \rho_{xy}\simeq -{h\over \t}\,,
\label{ss6}\ee
and the inverse Hall angle becomes
\be
\cot \Theta_H\equiv {\s_{xx}\over \s_{xy}}\simeq {C\over Z_2~h}\sqrt{1+{Z_1^2Z_2^4\over \t^2}h^2}={C\over Z_2~h}+{\cal O}(h)\,.
\label{ss8}
\ee

\item The Charge-conjugate regime (CCR) when $Z_1Z_2~C\gg |\t|$ with conductivity given by
\be
  \s_{xx}\simeq Z_1Z_2^2{C\over \sqrt{C^2+Z_2^2~h^2}}\sp \s_{xy}={Z_2^2\t\over C^2+Z_2^2h^2}h\,,
\label{ss5}  \ee
\be
\rho_{xx}\simeq {\sqrt{C^2+Z_2^2~h^2}\over Z_1Z_2^2C}\sp \rho_{xy}\simeq -{\t ~h\over Z_1^2Z_2^2C^2}\,,
\label{ss7}\ee
\be
\cot\Theta_H\simeq {Z_1C\over \t~h}\sqrt{C^2+Z_2^2h^2}={Z_1C^2\over \t~h}+{\cal O}(h)\,.
\label{ss9}
\ee
\end{itemize}

It has been known for some time that the formulae above cannot describe the  passage between linear and quadratic resistivity seen in cuprates while keeping always the inverse Hall angle $\cot\Theta_H\sim T^2$~\cite{kiki}.
This can be seen from the formulae above: In the DR  from (\ref{ss4}) and (\ref{ss6}) we have $\rho_{xx}\sim \cot\Theta_H$ so it can accommodate the Fermi liquid behavior if $C/Z_2\sim T^2$.
In the CCR the correct Hall angle implies $C^2Z_1\sim T^2$. Using this and the previous relation we obtain for the resistivity in this regime $\rho_{xx}\sim {1\over Z_1Z_2^2}\sim T^2$ instead of linear resistivity.

In the presence of a non-trivial intrinsic T-violation ($S\not=0$), the conductivity formulae in the DR and CCR regimes (\ref{sss2}) and (\ref{sss3}) are modified as follows.
\begin{itemize}
\item In DR regime
\be
  \s_{xx}\simeq {Z_2\over C}|\t|\sp \sigma_{xy}=S\,.
\label{ss10}
 \ee
For the resistivity however we obtain
\be
\rho_{xx}\simeq {CZ_2|\t|\over \t^2Z_2^2+C^2S^2}\sp \rho_{xy}\simeq -{C^2S\over \t^2Z_2^2+C^2S^2}\,,
\label{ss12}
\ee
and
\be
\cot \Theta_H\simeq {Z_2|\t|\over CS}\,.
\label{ss13}
\ee

\item The Charge-conjugate regime (CCR) when $Z_1Z_2~C\gg |\t|$ with  conductivity given by
\be
  \s_{xx}\simeq Z_1Z_2^2\sp \sigma_{xy}=S\,,
\label{ss11}
\ee
while the resistivity and the inverse Hall angle are
\be
\rho_{xx}\simeq {Z_1Z_2^2\over Z_1^2Z_2^4+S^2}\sp \rho_{xy}\simeq -\frac{S}{Z_1^2 Z_2^4+S^2}\sp \cot\Theta_{H}={Z_1Z_2^2\over S}\,.
\ee

\end{itemize}

When $S\not =0$ there is a third scaling regime at zero magnetic field, beyond the two regimes DR and CCR discussed above. It appears when $|S|\gg Z_1Z_2^2$ and $C |S|\gg |\t| Z_2$. The leading resistivity in that case becomes
\be
\rho_{xx}\simeq {Z_2\over CS^2}\sqrt{\t^2+C^2Z_1^2Z_2^2}~~\simeq ~~     \left\{ \begin{array}{lll}
\displaystyle ~~{{Z_2|\t|}\over CS^2}\,,&\phantom{aa} &DR~~,\\ \\
\displaystyle ~~{Z_1Z_2^2\over S^2}\,,&\phantom{aa}&CCR\,.
\end{array}\right.
\ee

\section{Axion-Driven Hyperscaling-Violating Geometry without a Charge Sector}\label{HSV}
In this section we show that a scaling  hyperscaling-violating geometry can be supported by scalars without a potential (aka axions) without the presence of  charge sector, i.e., bulk gauge field. The bulk theory we are considering is the Einstein-Dilaton-Axion theory.

The action reads
\begin{equation}\label{hsvaction}
S_1=M_p^2 \int d^{4}x\sqrt{-\det g} \left[\mathcal{R}-\frac{1}{2}(\partial\phi)^2+V(\phi)-\frac{Y(\phi)}{2}\sum_{I=1}^{2}(\partial \alpha_I)^2\right],
\end{equation}
where the axions $\alpha_I$ will be linear in $(x,y)$ coordinates and will  break translational symmetry and hence provide an effect of momentum dissipation or a sort of mean-field holographic disorder. We will be interested in IR scaling solutions and we will approximate the dilaton couplings $V$ and $Y$ in the IR with exponentials, i.e.,
\begin{equation}\label{asympvy }
V(\phi)\sim V_0 \,e^{-\delta\, \phi},\quad Y(\phi)\sim e^{\eta\,\phi}\,,
\end{equation}
with $\delta, \eta$ two constants.

We assume the (black) hyperscaling-violating (rotationally-symmetric) metric ansatz
\begin{eqnarray}\label{axionfinitT}
ds^2&=&r^\theta\left(-f(r)\frac{dt^2}{r^{2z}}+\frac{L^2 dr^2}{r^2 f(r)}+\frac{dx^2+dy^2}{r^2}\right)\,,\\
\phi&=&\phi_0+\kappa\log(r)\sp \alpha_1=k\,x\sp \alpha_2=k\,y\,,\nonumber
\end{eqnarray}
with $(L, \phi_0, \kappa, k)$ constants. We obtain the black brane solution
\begin{eqnarray}\label{solutionhsv}
&& f(r)=1-\left(\frac{r}{r_0}\right)^{2+z-\theta}\sp z=\frac{\eta^2-\delta^2+1}{\eta(\eta+\delta)}\sp \theta=-\frac{2\delta}{\eta},\quad \kappa=-\frac{2}{\eta}\,,\nonumber\\
&&L^2=\frac{(z+2-\theta)(2z-\theta)}{e^{-\delta\phi_0}V_0 }\sp e^{\eta\phi_0}k^2 L^2=2(z-1)(z+2-\theta)\,,\\
&& \delta=\pm\frac{\theta}{\sqrt{(\theta-2)(\theta+2-2z)}}\sp \eta=\mp\frac{2}{\sqrt{(\theta-2)(\theta+2-2z)}}\,.\nonumber
\end{eqnarray}
To have a well defined geometry and a resolvable  singularity, one should consider the Gubser criterion which restrict the parameter parameter of $(z, \theta)$ appearing in~\eqref{solutionhsv}. It turns out that the allowed parameter range is given by
\begin{eqnarray}
\text{IR}\quad r\rightarrow 0&:& [z\leqslant 0, \theta>2],\quad [0<z<1, \theta>z+2]\,,\label{airlocates0}\\
\text{IR}\quad r\rightarrow \infty&:& [1<z\leqslant 2, \theta<2z-2],\quad [z>2, \theta<2]\label{birlocates}\,.
\end{eqnarray}
A similar solution has been studied by the author of~\cite{Gouteraux:2014hca} with a different backeround ansatz.

\section{Hyperscaling-Violating Geometry with a Charge Sector}\label{HSVu1}
In this appendix we show that the hyperscaling-violating geometry~\eqref{fixpoint} can be realised by the Einstein-Maxwell-Dilaton (EMD) theory in presence of charge sector, i.e, bulk gauge field.

The bulk theory is given by
\begin{equation}\label{emdaction}
S_2=M_p^2\int dx^4\sqrt{-\det g}\left[\mathcal{R}-\frac{1}{2}(\partial\phi)^2-\frac{\hat{Z}(\phi)}{4}H^2+\hat{V}(\phi)\right],
\end{equation}
with $H_{\mu\nu}=\partial_\mu B_\nu-\partial_\nu B_\mu$. Assuming that in the IR $\hat{V}(\phi)$ and $\hat{Z}(\phi)$ have exponential asymptotics as in supergravity,
\begin{eqnarray}
\hat{V}(\phi) \sim V_0\, e^{-\delta \phi}\sp \hat{Z}(\phi)\sim \hat{Z}_0 e^{\gamma \phi}\,,
\label{c2}\end{eqnarray}
we can obtain the quantum critical  geometry driven by the running scalar~\cite{cgkkm}
\begin{eqnarray}\label{hyperscal}
&&ds^2=-\left(\frac{r}{\ell}\right)^{\theta-2z}f(r) dt^2+\left(\frac{r}{\ell}\right)^{\theta-2}\frac{L^2dr^2}{f(r)}+\left(\frac{r}{\ell}\right)^{\theta-2}(dx^2+dy^2)\,,\\  \nonumber
&&f(r)=1-\left(\frac{r}{r_0}\right)^{2+z-\theta},\; B=\sqrt{\frac{2(z-1)}{\hat{Z}_0 (z-\theta+2)}}\left(\frac{\ell}{r}\right)^{z-\theta+2}f(r) dt,\; e^{\phi}=\left(\frac{r}{\ell}\right)^{\sqrt{(\theta-2)(\theta-2z+2)}}\,,\\\nonumber
&&L^2=\frac{(z-\theta+1)(z-\theta+2)}{V_0\ell^2},\;   \gamma=\frac{4-\theta}{\sqrt{(\theta-2)(\theta-2z+2)}},\; \delta=\frac{\theta}{\sqrt{(\theta-2)(\theta-2z+2)}}\,,
\end{eqnarray}
with
\be
z=\frac{\gamma^2+2\gamma\delta-3\delta^2+4}{\gamma^2-\delta^2}\sp
\theta=\frac{4\delta}{\gamma+\delta}\,,
\label{zt}\ee
where $r_0$ is the horizon and $\ell$ is the hyperscaling-violating scale.

Comparing with the case~\eqref{axionfinitT} in appendix~\ref{HSV}, one finds that two theories share the same form of background, especially the blackness function.

The temperature as well as the entropy density is given by
\begin{eqnarray}\label{backT}
T=\frac{a}{\ell}\left(\frac{\ell}{r_0}\right)^z\sp S_{en}=\frac{1}{4 G_N}\left(\frac{r_0}{\ell}\right)^{\theta-2}\sp a=\frac{|z-\theta+2|}{4\pi L}\,,
\end{eqnarray}
with $G_N$ the Newton constant. Note that the thermal entropy scales like\;\footnote{There is a very special case $z=0$, where the temperature is independent of $r_0$. It is clear that the condition~\eqref{linearcond} can not be satisfied as $z=0$, and thus we do not consider this case below.}
\begin{eqnarray}
S_{en}\sim T^{\frac{2-\theta}{z}}\,,
\end{eqnarray}
which gives an interpretation to the hyperscaling violation exponent $\theta$.

The corresponding extreme IR fixed point is given by~\eqref{fixpoint} with the allowed parameter space depending on the location of the IR regime:
\begin{itemize}
  \item (a) $[z< 0, \theta>2]$: the extreme IR is at $r\rightarrow 0$ and  the  extremal solution has $T\rightarrow 0$.
  \item (b) $[0<z<1, \theta>z+2]$: the IR limit is at $r\rightarrow 0$ and  the solution has $T\rightarrow \infty$ in the near extremal geometry.
  \item (c) $[1<z\leqslant 2, \theta<2z-2]$ or $[z>2, \theta<2]$ : in the extremal limit the IR is at $r\rightarrow \infty$ and the corresponding solution has $T\rightarrow 0$.
\end{itemize}
Such hyperscaling-violating geometries describe the quantum criticality at fractionalized phases as in the deep IR the degrees of freedom ``behind" the extremal horizon generate a non-zero flux associated with $B_\mu$.

\section{DBI Stress Tensor and Probe Limit}\label{probecondition}
Since we are working in the probe DBI limit, the contribution of the DBI part should be subleading to the gravity background. The geometry should be provided by other parts of matter content. The full theory may be written as
\begin{eqnarray}\label{fullaction}
S=S_g+S_{DBI}, \quad S_g=M_p^2 \int dx^4\sqrt{-\det g}\,\left[\mathcal{R}+\mathcal{L}_m\right],
\end{eqnarray}
where $\mathcal{L}_m$ is used to seed the environment~\eqref{backgd}.

Varying the action~\eqref{dbiaction}, we obtain the DBI stress tensor
\be
T^{DBI}_{\m\n}=-\frac{1}{\sqrt{-\det g}}\frac{\delta S_{DBI}}{\delta g^{\mu\nu}}=-{1\over 2}Z_1\sqrt{\frac{\det (g+  Z_2 F)}{\det g}}g_{\m\r}(g+ Z_2 F)^{\r\s}_sg_{\s\n}\,,
\ee
where the subscript $``s"$ means the symmetric part and $(g+ Z_2 F)^{\r\s}$ is the inverse of $(g+ Z_2 F)_{\r\s}$. The full Einstein equations are given by
\begin{eqnarray}
G_{\mu\nu}=\mathcal{R}_{\mu\nu}-\frac{1}{2}\mathcal{R} g_{\mu\nu}=T^m_{\mu\nu}+T^{DBI}_{\mu\nu}\,,
\end{eqnarray}
with $T^m_{\mu\nu}$ the stress tensor from $\mathcal{L}_m$.
To derive the condition for the probe description, we must require the stress tensor of DBI action should be much smaller than the contribution from other matter contents. As a consequence, one must require that
\begin{equation}
T_{\mu\nu}^{DBI}\ll G_{\mu\nu}\sim T^m_{\mu\nu}\,.
\end{equation}

The dual geometry for quantum criticality~\eqref{fixpoint} as well as~\eqref{fixpointT} can be obtained from many setups and two examples are presented in appendix~\ref{HSV} and appendix~\ref{HSVu1}. The geometry~\eqref{axionfinitT} or~\eqref{hyperscal} can be considered as the near horizon limit of the full background~\eqref{backgd} that interpolates between $AdS_4$ regime in the UV and hyperscaling-violating regime in the IR.

Notice that the DC conductivity is fully computed in the IR region, i.e., $r_s\rightarrow r_0$ and two theories in appendices~\ref{HSV} and~\ref{HSVu1} share the same form of metric. Here we focus on the IR limit~\eqref{hyperscal}, but our discussion below is also valid for the case in appendix~\ref{HSV}.

We work out the condition for the probe description with $E\rightarrow0$ as required by the linear approximation. One can see from~\eqref{jxex} and~\eqref{jyex} that $(\x, \y)\rightarrow0$ as $E\rightarrow0$.
So we finally obtain that
\begin{eqnarray}\label{finitetem}
\nonumber G_{tt} &=&\frac{f(2-\theta)}{4 L^2 \ell^{2-2z} r^{2z}}\left[\theta-6+(\theta+2-2z)\left(\frac{r}{r_0}\right)^{2+z-\theta}\right]\,,\\
G_{rr} &=&\frac{(\theta-2)}{4 f r^2}\left[3\theta-2-4z+(2z-2-\theta)\left(\frac{r}{r_0}\right)^{2+z-\theta}\right]\,,\\
\nonumber G_{xx}&=&G_{yy} =\frac{1}{4L^2 r^2}\left[4z^2+(\theta-2)^2-4z(\theta-1)+(\theta-2)(\theta+2-2z)\left(\frac{r}{r_0}\right)^{2+z-\theta}\right]\,,
\end{eqnarray}
and
\begin{eqnarray}
T_{tt}^{DBI}&=&\frac{f \Omega {\ell}^{2z} }{2r^{2z} Z_2^2}\sp T_{rr}^{DBI} =-\frac{L^2\Omega \ell^2}{2f r^2 Z_2^2}\,,\\
T_{xx}^{DBI}&=&T_{yy}^{DBI}=-\frac{r^{2\theta}Z_1^2 Z_2^2 \ell^{2-2\theta}}{2 r^{2} \Omega}\,,
\end{eqnarray}
with all non-diagonal terms zero. Here we have defined
\begin{equation}\label{oga}
\Omega=Z_2\sqrt{[(\t-h S)^2+h^2 Z_1^2 Z_2^4] (r/\ell)^4+Z_1^2 Z_2^2\, (r/\ell)^{2\theta}}\,.
\end{equation}

Note that we are working in the near horizon limit $r\rightarrow r_0$. Therefore, the terms in the brackets of~\eqref{finitetem} are in general at order one.\footnote{In the special case $\theta=2$, $G_{tt}$ and $G_{rr}$ vanish. In this case we should demand that $T_{tt}^{DBI}\ll M_p^{2},T_{rr}^{DBI}\ll M_p^{2}$. However, one can see that $T_{rr}^{DBI}$ is divergent at the horizon. Therefore, the case $\theta=2$ should be excluded.} The probe approximation demands $T_{\mu\nu}^{DBI}\ll M_p^2 G_{\mu\nu}$, i.e.,
\begin{equation}
\Omega\ll Z_2^2 M_p^2/\ell^2\sp  (r/\ell)^{2\theta}\,Z_1^2 Z_2^2 \ell^2 M_p^{-2}\ll \Omega\,,
\end{equation}
which results in
\begin{equation}\label{probe}
Z_1(r_0)^2Z_2(r_0)^2 \left(\frac{r_0}{\ell}\right)^{2\theta}\frac{\ell^2}{M_p^{2}} \ll \Omega(r_0) \ll \frac{Z_2(r_0)^2 M_p^2}{\ell^2}\,,
\end{equation}
evaluated at the horizon $r_0$.

Note that $Z_1$ and $Z_2$ in general follow a power function of $r$  in the far IR region $r\rightarrow r_0$, therefore are temperature dependent.
We assume that $Z_1$ and $Z_2$ have the following form in the far IR:
\begin{equation}
Z_1=\textbf{A}\left(\frac{r}{\ell}\right)^\alpha, \quad Z_2= \textbf{B}\left(\frac{r}{\ell}\right)^\beta,
\end{equation}
which equivalently means that $Z_1$ and $Z_2$ behave as $e^{k\phi}$ with $k$ a constant.
Note that $C=(r/\ell)^{\theta-2}$ and $T=a(\ell/r_0)^z/\ell$ as seen from~\eqref{backT}. From now on, we focus on the case of interest~\eqref{linearcond}. In order to satisfy the condition~\eqref{linearcond}, $Z_1$ and $Z_2$ can be in general given by
\begin{equation}\label{linearZ}
Z_1=\frac{Z_0 a}{\ell\ell_0^2}\left(\frac{r}{\ell}\right)^{4-z-2\theta}, \quad Z_2=\frac{\ell\ell_0}{a}\left(\frac{r}{\ell}\right)^{z+\theta-2},
\end{equation}
where $Z_0$ and $\ell_0$ are defined in~\eqref{linearcond}. By using~\eqref{linearZ}, the constraint~\eqref{probe} can be expressed in terms of temperature.

The constraint~\eqref{probe} can be rewritten by using dimensionless variables~\eqref{dimenless},
\begin{equation}\label{probenew}
N_\alpha\textbf{T}^{\frac{\theta+z-6}{z}}\ll\textbf{T}^{-\frac{2}{z}}\sqrt{1+\tau^2 +\textbf{h}^2/\textbf{T}^2}\ll N_\beta\textbf{T}^{\frac{2-\theta-z}{z}}\,,
\end{equation}
with $N_\alpha=\frac{a|\tau|Z_0\ell}{M_p^2\ell_0^2}\left(\frac{Z_0\ell}{a}\right)^{\frac{z+\theta-4}{z}}$ and $N_\beta=\frac{|\tau|M_p^2\ell_0^2}{aZ_0\ell}\left(\frac{Z_0\ell}{a}\right)^{\frac{4-z-\theta}{z}}$ two dimensionless quantities. The above constraint depends on the sign of $(\theta+z-4)/z$.

\subsection{Low temperature case $(\theta+z-4)/z>0$ }
As $(\theta+z-4)/z>0$, by comparing the left and the right parts of~\eqref{probenew} one can immediately find that $\textbf{T}\ll 1$. This corresponds to the low temperature limit. As the constraint depends on the magnetic field, we must explore two cases.

 \begin{itemize}
\item (E-1a) If $\textbf{h}/\textbf{T}$ is at order one or even smaller, we find that~\eqref{probenew} is satisfied automatically when $(\theta+z-4)/z>0$.
\item (E-1b) In the second case $\textbf{h}/\textbf{T}\gg 1$, one obtains from~\eqref{probenew} that
\begin{equation}\label{applowth}
\textbf{h}\ll\textbf{T}^{\frac{4-\theta}{z}}\,.
\end{equation}
Note that $\textbf{T}\ll 1$ and $(4-\theta)/z<1$.
\end{itemize}
Therefore, in the low temperature case $(\theta+z-4)/z>0$ the magnetic field can not be too strong and the upper bound is given by~\eqref{applowth}.

\subsection{High temperature case $(\theta+z-4)/z<0$ }
As $(\theta+z-4)/z<0$,  from~\eqref{probenew} one finds that $\textbf{T}\gg 1$, i.e., the high temperature limit. Similarly, we also need to consider two cases.
\begin{itemize}
\item (E-2a) If $\textbf{h}/\textbf{T}$ is at order one or even smaller, the constraint~\eqref{probenew} is automatically satisfied in this case.

\item (E-2b) On the other hand, if $\textbf{h}/\textbf{T}\gg 1$, we can obtain that
\begin{equation}\label{hightllh}
\textbf{h}\ll \textbf{T}^{\frac{4-\theta}{z}}\,.
\end{equation}
The above constraint looks the same as~\eqref{applowth}. However, note that $\textbf{T}\gg 1$ and $(4-\theta)/z>1$ in the present case.
\end{itemize}
Therefore, in the high temperature case $(\theta+z-4)/z<0$ we obtain a similar bound for the magnetic field.

\section{Higher Derivative Corrections}\label{highorder}
Even at classical level, there are in principle higher derivative corrections $\mathcal{O}(\partial F)$ in the DBI action~\eqref{dbiaction}. We hope such corrections are small, and therefore can not change our result. The lowest corrections can have the form,
\begin{equation}
L_1=\nabla_\lambda F_{\mu\nu} \nabla^\lambda F^{\mu\nu}\sp \text{or}\quad L_2=F_{\mu\nu}\nabla_\lambda\nabla^\lambda F^{\mu\nu}\,,
\end{equation}
which should be much smaller than the leading order term $\ell_s^{-2} F_{\mu\nu} F^{\mu\nu}$ with $\ell_s$ the string length.

We compute those terms by using the same background as in appendix~\ref{probecondition}. We also work in the limit $E\rightarrow0$ and take $S=0$, then we obtain that
\begin{equation}
\ell_s^{-2}F_{\mu\nu} F^{\mu\nu}=\frac{2 r^4}{\ell_s^2\ell^4}\left(\frac{\ell^{2\theta} h^2}{r^{2\theta}}-\frac{\t^2}{\Omega^2}\right)\,,
\end{equation}
\begin{equation}
L_1=\frac{f r^{4-\theta}}{L^2\ell^{6-\theta}}\left[\frac{3(\theta-2)^2h^2\ell^{2\theta}}{r^{2\theta}}-\frac{\t^2}{\Omega^2}\left(\theta^2-4\theta+12-\frac{8r\Omega'}{\Omega}+\frac{2r^2\Omega'^2}{\Omega^2}\right)\right]\,,
\end{equation}
\begin{equation}
L_2=-\frac{f' r^{5-\theta}}{L^2\ell^{6-\theta}}\left[\frac{2(\theta-2)\ell^{2\theta}h^2}{r^{2\theta}}+\frac{4\t^2}{\Omega^2}-\frac{2\t^2 r \Omega'}{\Omega^3}\right]+\frac{f r^{4-\theta}}{L^2\ell^{6-\theta}}\left[\frac{(\theta-2)(2+2z-\theta)\ell^{2\theta}h^2}{r^{2\theta}}\right.
\end{equation}
$$
\left.+\frac{\t^2}{\Omega^2}\left(4+4z-8\theta+\theta^2+\frac{2r((3-z+\theta)\Omega'+r\Omega'')}{\Omega}-\frac{4 r^2 \Omega'^2}{\Omega^2}\right)\right],
$$
where $\Omega$ is defined in~\eqref{oga} with $S=0$.

Remember that we are working in the far IR regime, the above results are valid for the near horizon geometry~\eqref{hyperscal}. Since the blackness factor  $f(r)\rightarrow 0$ in the IR region, $L_1$ and the second part of $L_2$ just vanish at the horizon. By using the condition~\eqref{linearcond} and the form of $f(r)$ in~\eqref{hyperscal}, we evaluate them at the horizon and convert $r_0$ into temperature. We then obtain
\begin{equation}
\ell_s^{-2}F_{\mu\nu} F^{\mu\nu}|_{r=r_0}=\frac{2\t^2}{\ell_s^{2}} \left(\frac{Z_0\ell}{a}\right)^{\frac{2\theta-4}{z}} \textbf{T}^{\frac{2\theta-4}{z}}\left[ \textbf{h}^2-\frac{\textbf{T}^4\tau^2}{\textbf{h}^2+(1+\tau^2)\textbf{T}^2}\right]\,,
\end{equation}
\begin{eqnarray}
\nonumber L_2|_{r=r_0}&&=\frac{2\t^2(2+z-\theta)}{L^2\ell^2} \left(\frac{Z_0\ell}{a}\right)^{\frac{3\theta-4}{z}}\times\\
&&\textbf{T}^{\frac{3\theta-4}{z}}\left[(\theta-2)\textbf{h}^2-\frac{\textbf{h}^2\textbf{T}^4(\theta+2 z-2)\tau^2+\textbf{T}^6(\theta+z-2)\tau^2(1+\tau^2)}{(\textbf{h}^2+(1+\tau^2)\textbf{T}^2)^2}\right]\,.
\end{eqnarray}
To obtain above expressions we have used the relation~\eqref{linearZ}.

Note that $(2+z-\theta)$ can not vanish because of the Gubser criterion. By demanding $ L_2\ll \ell_s^{-2}F_{\mu\nu} F^{\mu\nu}$, we arrive at the result
\begin{eqnarray}\label{constderiv}
\nonumber N_0\textbf{T}^{-\frac{\theta}{z}}&&\left|\textbf{h}^2-\frac{\textbf{T}^4\tau^2}{\textbf{h}^2+(1+\tau^2)\textbf{T}^2}\right|\gg\\
&&\left|(\theta-2)\textbf{h}^2-\frac{\textbf{h}^2\textbf{T}^4(\theta+2 z-2)\tau^2+\textbf{T}^6(\theta+z-2)\tau^2(1+\tau^2)}{(\textbf{h}^2+(1+\tau^2)\textbf{T}^2)^2}\right|\,,
\end{eqnarray}
with $N_0=\frac{L^2 \ell^2}{|2+z-\theta|\ell_s^2}\left(\frac{Z_0\ell}{a}\right)^{-\theta/z}$.
Notice that the parameter space with $\theta=2$ or $z=0$ has been excluded. To combine~\eqref{constderiv} with the result of appendix~\ref{probecondition} on back-reaction, we need to consider the low temperature limit $\textbf{T}\ll 1$ as well as the high temperature limit $\textbf{T}\gg 1$.

\subsection{Low temperature limit  $\textbf{T}\ll 1$}
In the low temperature limit  $\textbf{T}\ll 1$, the constraint~\eqref{constderiv}  also depends on the magnetic field, so we need to explore two cases.

\begin{itemize}
\item (F-1a) We first consider the case $\textbf{h}/\textbf{T}\ll 1$, the constraint~\eqref{constderiv} becomes
\begin{eqnarray}
&&N_0\textbf{T}^{-\frac{\theta}{z}}\gg  N_1|\theta+z-2|\sp (\theta+z-2)\neq 0\,,\label{lowh}\\
&&N_0\textbf{T}^{-\frac{\theta}{z}}\gg  N_2 \frac{\textbf{h}^2}{\textbf{T}^2}\sp\,\quad\quad\quad (\theta+z-2)= 0\,,\label{z2const}
\end{eqnarray}
where $N_1, N_2$ are two positive constants. The latter one~\eqref{z2const} corresponds to the theory with $Z_2$ a constant.
\item (F-1b) On the other hand, if $\textbf{h}\gg\textbf{T}$, the condition~\eqref{constderiv} can be reduced to
\begin{equation}\label{highh}
N_0\textbf{T}^{-\frac{\theta}{z}}\gg N_3\,,
\end{equation}
with $N_3$ a positive constant. This is also true when $\textbf{h}$ is of the same order as $\textbf{T}$.
\end{itemize}
Therefore, the suppression of higher derivative corrections in general demands
\begin{equation}
\theta/z>0\,,
\end{equation}
when $\textbf{T}\ll 1$. Note that in the special case~\eqref{z2const} with $(\theta+z-2)= 0$, one can also choose $\theta=0$.

\subsection{High temperature limit  $\textbf{T}\gg 1$}
We now turn to the high temperature case. The discussion is very similar as before.
\begin{itemize}
\item (F-2a)  When $\textbf{h}/\textbf{T}\ll 1$, the constraint~\eqref{constderiv} is reduced to~\eqref{lowh} as $(\theta+z-2)\neq 0$ and~\eqref{z2const} as $(\theta+z-2)= 0$.
\item (F-2b) In contrast, if $\textbf{h}/\textbf{T}\gg 1$, we obtain~\eqref{highh} but with $\textbf{T}\gg1$.
\end{itemize}
Therefore, in the high temperature case $\textbf{T}\gg 1$, the constraint from the higher derivative corrections in general determines
\begin{equation}
\theta/z<0\,,
\end{equation}
and in very spacial case one can choose $\theta=0$.

\section{Scaling Solutions with Two Scalars}\label{app2scalars}

We consider an effective gravitational action of the form
\be
S=M_p^2\int d^4x\sqrt{-\det g}\left[\mathcal{R}-{1\over 2}(\pa\vec \phi)^2+V_0~e^{-\vec \d\cdot \vec \phi}-{1\over 4}\sum_{I=1}^N~e^{\vec \g_I\cdot \vec \phi}F_I^2\right],
\label{g0}\ee
which contains apart from the metric two scalars and $N$ massless vectors. We use a vector notation with $\vec \phi\sim \phi^i\sim (\phi_1,\phi_2)$. The action depends on $N+1$ 2-vectors $\vec \d$ and $\vec \g_I$.

We now make the scaling ansatz~\footnote{The scaling solutions in a more general theory with an arbitrary number of scalars and vector fields are discussed in~\cite{Li:2016rcv}. Please consult~\cite{Li:2016rcv} for more details.}
\be\label{hsv2scalars}
ds^2=r^{\theta}\left[-{dt^2\over r^{2z}}+{L^2\,dr^2+d\vec x^2\over r^2}\right]\sp \vec\phi=\vec \kappa\,\log r\sp A_{tI}= A_{tI}(r)\,,
\ee
with $I=1,\dots, N$.
Solving equations for the vectors we can determine,
\be\label{At}
A_{tI}=\mu_I+Q_I~ r^{2-z-\vec \g_I\cdot \vec \kappa}\sp 2-z-\vec \g_I\cdot \vec \kappa\not=0\,.
\ee
Here $\mu_I$, $Q_I$ are constants and $\vec\kappa$ is a constant 2-vector. Note that when $\vec\g_I\cdot \vec\kappa=2-z$, we instead obtain a logarithmically running gauge field.
We do not consider this kind of solution.

\subsection{$N=1$ case}
We start by solving the $N=1$ case. In this case the Einstein equations give
\be
\vec\kappa \cdot \vec \g_1=4-\theta\sp \vec\kappa \cdot \vec \d=\theta\sp \vec \kappa^2=(\theta-2)(\theta+2-2z)\,,
\label{g1}\ee
\be
Q_1^2={2(z-1)\over z+2-\theta}\sp L^2=\frac{(z-\theta+1)(z-\theta+2)}{V_0}\,,
\label{g2}
\ee
while the two scalar equations
\be
\g^1_1~Q_1^2(z-\theta+2)^2-2\d _1~L^2\,V_0-2\kappa_1~(z-\theta+2)=0\,,
\label{g3}\ee
\be
\g^2_1~Q_1^2(z-\theta+2)^2-2\d_2~L^2\,V_0-2\kappa_2~(z-\theta+2)=0\,,
\label{g4}\ee
where we have partly used equations (\ref{g1}). Using also (\ref{g2}) they become
\be
 \left[(z-1)~\g^1_1-\d_1~(z-\theta+1)-\kappa_1\right]~(z-\theta+2)=0\,,
 \label{g5}\ee
 \be
 \left[(z-1)~\g^2_1-\d_2~(z-\theta+1)-\kappa_2\right]~(z-\theta+2)=0\,.
 \label{g6}
 \ee
 We assume further that $z-\theta+2\not=0$ as otherwise the potential must be subleading and we obtain a logarithmically running gauge field.  Then~\eqref{g5} and~\eqref{g6} become
\be
 \vec \kappa=(z-1)~\vec \g_1-(z-\theta+1)~\vec \d\,.
 \label{g7}
 \ee
 Using this in~\eqref{g1} we can compute
 \be
 \theta=2{   2\vec \g_1 \cdot \vec \d- 2\vec \d^2+(\vec \g_1\cdot \vec \d)^2-\vec \g_1^2\vec \d^2\over  \vec \g_1^2-\vec \d^2+(\vec \g_1\cdot \vec \d)^2-\vec \g_1^2\vec \d^2}\,,
\label{g8}\ee
\be
z={-3\vec\d^2+2\vec \g_1\cdot \vec \d+\vec\g_1^2+(\vec \g_1\cdot \vec \d)^2-\vec \g_1^2\vec \d^2+4\over  \vec \g_1^2-\vec \d^2+(\vec \g_1\cdot \vec \d)^2-\vec \g_1^2\vec \d^2}\,.
\label{g9}\ee
Then~\eqref{g2},~\eqref{g7},~\eqref{g8} and~\eqref{g9} give the complete solution.
We can also invert and write
\be
\vec \d^2={\theta-(z-1)\vec \g_1\cdot \vec \d\over \theta-z-1}\sp \vec \g_1^2={4-\theta+(z+1-\theta)\vec \g_1\cdot \vec \d\over z-1}\,.
\label{g10}\ee

Consider now an orthonormal  basis in the two-dimensional space with basis vectors
\be
\vec e_1={(z-1)~\vec \g_1-(z-\theta+1)~\vec \d\over \sqrt{(\theta-2)(\theta+2-2z)}}\,,
\label{g11}\ee
\be
\vec e_2={\sqrt{(z-1)(\theta-z-1)}(\theta \vec \g_1+(\theta-4)\vec \d)\over \sqrt{(2-\theta)(\theta+2-2z)(\theta(\theta-4)+(\theta-2)(\theta+2-2z)\vec \g_1\cdot \vec \d)}}\,,
\label{g111}\ee
with
\be
\vec e_1\cdot \vec e_1=\vec e_2\cdot \vec e_2=1\sp \vec e_1\cdot \vec e_2=0\,.
\label{g12}\ee
If we now write
\be
\vec \phi=\Phi_1\,\vec e_1+\Phi_2\,\vec e_2\,,
\label{g121}
\ee
then the action in (\ref{g1}) can be written as
\be
S=M_p^2\int d^4x\sqrt{-\det g}\left[\mathcal{R}-{1\over 2}(\pa\vec \Phi)^2+V_0~e^{-\vec \Delta\cdot \vec \Phi}-{1\over 4}~e^{\vec \Gamma_1\cdot \vec \Phi}F_1^2\right]\,,
\label{g00}\ee
where
\be
\vec \Delta=\left({\theta\over \sqrt{(\theta-2)(\theta+2-2z)}},\sqrt{(z-1)(\theta(\theta-4)+(\theta-2)(\theta+2-2z)\vec \g_1\cdot \vec \d)\over (\theta-2)(\theta+2-2z)(z+1-\theta)}\right)\,,
\label{g13}\ee
\be
\vec \Gamma_1=\left({4-\theta\over \sqrt{(\theta-2)(\theta+2-2z)}},\sqrt{(z+1-\theta)(\theta(\theta-4)+(\theta-2)(\theta+2-2z)\vec \g_1\cdot \vec \d)\over (\theta-2)(\theta+2-2z)(z-1)}\right)\,,
\label{g14}\ee
with
\be
\vec \Delta^2=\vec \d^2\sp \vec \Gamma_1^2=\vec \g_1^2\sp \vec \Delta\cdot \vec \Gamma_1=\vec \delta\cdot \vec \gamma_1\,.
\label{g15}\ee
In this new frame the scalar solution is
\be
\Phi_1=\sqrt{(\theta-2)(\theta+2-2z)}~\log r\sp \Phi_2=0\,.
\label{g16}\ee

\subsection{Case with several vector fields}
We now consider the general case with several gauge fields.
In order for them to contribute at the same order, there are two possibilities for each one:
\begin{enumerate}

\item $\vec\g_I\cdot \vec\kappa=4-\theta$.

\item $\vec\g_I\cdot \vec\kappa=2-z$. This seeming solution is a fake, because in this case we obtain a logarithmically running gauge field.

\end{enumerate}

We will consider all gauge fields that are in the first case. Then we have
\be
\vec\kappa \cdot \vec \g_I=4-\theta, \forall I\sp \vec\kappa \cdot \vec \d=\theta\sp \vec \kappa^2=(\theta-2)(\theta+2-2z)\,,
\label{g17}\ee
\be
\sum_I Q_I^2={2(z-1)\over z+2-\theta}\sp L^2=\frac{(z-\theta+1)(z-\theta+2)}{V_0}\,,
\label{g18}\ee
and from the scalar equations
\be
\vec \kappa =(\theta-z-1)\vec\delta+{1\over 2}(z+2-\theta)\sum_I Q_I^2~\vec \g_I\,.
\label{g19}\ee
Substituting~\eqref{g19} into~\eqref{g17} we obtain the following conditions that determine (in principle) the $Q_I$ and $z,\theta$.
\be
(\theta-z-1)\vec\gamma_I\cdot\vec\d+{z+2-\theta\over 2}\sum_J \vec\g_I\cdot\vec\g_J ~Q_J^2=4-\theta\sp \forall ~I
\label{g20}\ee
\be
(\theta-z-1)\vec\d^2+{z+2-\theta\over 2}\sum_I \vec\g_I\cdot\vec\d ~Q_I^2=\theta\sp \sum_I Q_I^2={2(z-1)\over z+2-\theta}\,.
\label{g21}\ee

In order to solve these equations we define
\be
x_{IJ}\equiv \vec \g_I\cdot\vec\g_J\sp x_I\equiv \vec \g_I\cdot\vec\d\sp x_0\equiv \vec \d^2\,,
\label{g41}\ee
and assume that the inverse of the matrix $x_{IJ}$ exists.
So we can first solve the first set of equations~\eqref{g20} to determine $Q_I$.
\be
Q_I^2=2{(4-\theta)\over z+2-\theta}\sum_{J}(x^{-1})_{IJ}+2 {z+1-\theta\over z+2-\theta}\sum_{J}(x^{-1})_{IJ}x_J\,.
\label{g42}\ee
Then the two equations (\ref{g21}) become
\be
(4-\theta)\sum_{I,J}(x^{-1})_{IJ}x_J+(z+1-\theta)\left[\sum_{I,J}x_I(x^{-1})_{IJ}x_J-x_0\right]=\theta\,,
\label{g43}
\ee
\be
(4-\theta)\sum_{I,J}(x^{-1})_{IJ}+(z+1-\theta)\sum_{I,J}(x^{-1})_{IJ}x_J=(z-1)\,.
\label{g44}
\ee
We further introduce the scalar combinations
\be
W_0\equiv \sum_{I,J}(x^{-1})_{IJ}\sp W_1\equiv \sum_{I,J}(x^{-1})_{IJ}x_J\sp W_2\equiv \sum_{I,J}x_I(x^{-1})_{IJ}x_J\,.
\label{g45}
\ee
Then $(z,\theta)$ can be solved from~\eqref{g43} and~\eqref{g44} as
 \be
z={1+4W_0+2W_1+W_2-3W_1^2+3W_0W_2-x_0(1+3W_0)\over
1-W_1^2+(1+W_0)W_2-x_0(1+W_0)}\,,
\label{g46}
\ee
\be
\theta=2{2W_1(1-W_1)+(1+2W_0)W_2-x_0(1+2W_0)\over
1-W_1^2+(1+W_0)W_2-x_0(1+W_0)}\,.
\label{g47}\ee

For this solution to exist, $x_{IJ}$ must have an inverse and the denominators in~\eqref{g46} and~\eqref{g47} must be non-vanishing.
If the denominators vanish there is on solution for $(z,\theta)$.
If the matrix $x_{IJ}$ has zero eigenvectors, the story will change. For each zero eigenvector $\xi_I$ we obtain  an  equation on $(z,\theta)$
alone that does not involve the charges:
\be\label{zeroeigen}
(z+1-\theta)\sum_I x_I\xi_I=(\theta-4)\sum_I\xi_I\,.
\ee
In the case of a single zero eigenvector, nothing special happens in that the solution remains unique.
The equation above is one of the equations constraining $(z,\theta)$.
If there are two or more zero eigenvectors, unless the ratios $\sum_I x_I\xi_I\over \sum_I\xi_I$
are the same for all such zero eigenvectors there is no solution except possibly $\theta=4, z=3$.
Notice that such value gives a zero $\vec\kappa^2$ as seen from~\eqref{g17}. This is not consistent with other conditions.

In another degenerate case where $\vec\delta=\vec\gamma_I$, we find  the first equation in~\eqref{g21}
has the left hand side identical as the left hand side of one of the equations in~\eqref{g20}.
In such a case from the right hand sides we obtain $\theta=2$. A solution further exists if the other equations are compatible with this value.

The general algorithm is now clear. The vectors $\vec\gamma_I$ must be split into two groups $(\vec\gamma_i,\vec\gamma_a)$.
In all cases the following equations are satisfied:
\be
 \vec\kappa \cdot \vec \d=\theta\sp \vec \kappa^2=(\theta-2)(\theta+2-2z)\sp L^2=\frac{(z-\theta+1)(z-\theta+2)}{V_0}\,.
\label{g29}\ee

The vectors of the first group, $\vec\gamma_i$ correspond to nontrivial charge densities that affect the solution.
They must satisfy
\be
\vec\kappa \cdot \vec \g_i=4-\theta, \forall ~i\sp \sum_i Q_i^2={2(z-1)\over z+2-\theta}\,,
\label{g30}\ee
and
\be
\vec \kappa =(\theta-z-1)\vec\delta+{1\over 2}(z+2-\theta)\sum_i Q_i^2~\vec \g_i\,.
\label{g31}\ee
These equations generically completely determine the parameters of the solution, $z,\theta,\vec\kappa, B_0, Q_i$.

The other group $\vec\gamma_a$ corresponds to the case in which the associated charge densities are subleading compared for the first group.
For this to happen, if we consider the $r\to\infty$ part of the geometry then
\be
\vec\g_a\cdot \vec \kappa<4-\theta \sp \forall ~~a
\ee
while the opposite inequality must hold in the $r\to 0$ part of the geometry.

\section{D-brane Dilaton Couplings\label{string}}

In this appendix we will explore the simplest  top-down setup related to our DBI scaling solutions.

We  consider a D-dimensional closed string action in the string frame~\cite{book}, as well as a $p+1$-dimensional D-brane action,
\be
S_{closed}=\int d^Dx\sqrt{g_{\s}}~e^{-2\Phi}~\left[{\cal R}+4(\nabla \Phi)^2-{H^2\over 12}-{e^{2\Phi}\over 4}F_C^2+V_0+\cdots\right]\,,
\ee
\be
S_{D}=\int d^{p+1}x~e^{-\Phi}~\sqrt{\det(g_{\s}+F)}\,,
\ee
where $F_C=dC$ is a RR field strength while $H=dB_2$ is the field strength of the NS two-form. Here we demand the integer $D\geqslant 5$.

We compactify both to 4 dimensions on a manifold with linear dimension $R$. After we introduce
\be
 e^{-\chi}=R^{D-4}~e^{-2\Phi}\,,
\ee
we obtain
\be
S_{closed}=\int d^4x\sqrt{g_{\s}}~e^{-\chi}~\left[{\cal R}+(\nabla \chi)^2-{(\nabla R)^2\over R^2}-{R^2\over 4}F_g^2-{1\over 4}F_B^2-{e^{\chi}R^{D-4}\over 4}F_C^2+V_0+\cdots\right]\,,
\ee
\be S_{D}=\int d^{p+1}x~R^{p-3}~e^{-\Phi}~\sqrt{\det(g_{\s}+~F)}=
\int d^4 x~R^{p-3-{D-4\over 2}}~e^{-{\chi\over 2}}~\sqrt{\det(g_{\s}+~F)}\,.
\ee
We map this to the Einstein frame
\be
g_{\s}=e^{\chi}~g_{E}\,,
\ee
so that the actions become
\be
S_{closed}=\int d^4x\sqrt{g_{E}}~\left[{\cal R}_E-{1\over 2}(\nabla \chi)^2-{(\nabla R)^2\over R^2}-{R^2e^{-\chi}\over 4}F_g^2-{e^{-\chi}\over 4}F_B^2-{R^{D-4}\over 4}F_C^2+V_0 e^{\chi}+\cdots\right]\,,
\label{stri}
\ee
\be
S_{D}=\int d^4 x~R^{p-3-{D-4\over 2}}~e^{{3\over 2}\chi}~\sqrt{\det(g_{E}+e^{-\chi}~F)}\,.
\ee

We define
\be
\phi_1={1\over \sqrt{3}}(-\chi+2\log{R})\sp \phi_2=\sqrt{2\over 3}(\chi+\log R)\,,
\ee
\be
\chi={\sqrt{2}\phi_2-\phi_1\over \sqrt{3}}\sp \log R={\sqrt{2}\phi_1+\phi_2\over \sqrt{6}}\,,
\ee
 then the two actions can be written as
\be
\nonumber S_{closed}=\int d^4x\sqrt{g_{E}}~~\left[{\cal R}_E-{1\over 2}\left[(\pa\phi_1)^2+(\pa\phi_2)^2\right]-{e^{\sqrt{3}\phi_1}\over 4}F_g^2-{e^{{\phi_1-\sqrt{2}\phi_2\over \sqrt{3}}}\over 4}F_B^2-\right.
\ee
\be
\left.-{e^{(D-4){\sqrt{2}\phi_1+\phi_2\over \sqrt{6}}}\over 4}F_C^2+V_0~e^{{-\phi_1+\sqrt{2}\phi_2\over \sqrt{3}}}+\cdots\right]\,,
\label{g40}
\ee
\be
S_{D}=\int d^4 x~e^{-{(5+D-2p)\over 2\sqrt{3}}\phi_1{+\sqrt{2}{(4-D+2p)\over 4\sqrt{3}}}\phi_2}~\sqrt{\det(g_{E}+e^{{1\over \sqrt{3}}\phi_1-\sqrt{2\over 3}\phi_2}~F)}\,.
\ee
We arrive at the following coupling functions $Z_{1,2}$ after matching with our initial parametrization (\ref{dbiaction}),
\begin{equation}
\begin{split}
&Z_1=e^{-{(5+D-2p)\over 2\sqrt{3}}\phi_1+\sqrt{2}{(4-D+2p)\over 4\sqrt{3}}\phi_2}\sp Z_2=e^{{{1\over \sqrt{3}}\phi_1-\sqrt{2\over 3}\phi_2}}\,,\\
&Z_1Z_2^2=e^{-{(1+D-2p)\over 2\sqrt{3}}\phi_1{-\sqrt{2}{(4+D-2p)\over 4\sqrt{3}}\phi_2}}\,.
\end{split}
\end{equation}

\subsection{Scaling solutions}
For our case in~\eqref{g40}, we have a top-down theory with two scalars and three massless vectors. So we can apply the discussion of the previous appendix to find the corresponding hyperscaling violation geometries in our string setup. We denote charges associated with $F_g$, $F_B$ and $F_C$ as $Q_1$, $Q_2$ and $Q_3$, respectively, and read off the following vectors
\be\label{2vectors}
\vec\g_1=( \sqrt{3},0)\sp
\vec \d=\vec\g_2=\left({1\over  \sqrt{3}},-{\sqrt{2}\over \sqrt{3}}\right)\sp \vec\g_3=(D-4)\left({1\over  \sqrt{3}}, {1\over \sqrt{6}}\right)\,,
\ee
from the action~\eqref{g40}. Then we obtain
\be\label{Xij}
x_{IJ}=\left(\begin{matrix} 3, & 1, & D-4\\
1, & 1, & 0\\
D-4, &0,& {(D-4)^2\over 2}
\end{matrix}\right),
 \ee
 \be
 x_I=\left(1,  1,0\right)\sp x_0=1\,.
\ee
Moreover, we have the coupling functions in DBI
\be
Z_{1}\sim e^{\vec \a_{1}\cdot \vec\phi}\sp Z_{2}\sim e^{\vec \a_{2}\cdot \vec\phi}\,,
\ee
with
\be
\vec\alpha_1=\left(-{5+D-2p\over 2\sqrt{3}},{4-D+2p\over 2\sqrt{6}}\right)\sp \vec\alpha_2=\left({1\over \sqrt{3}},-{\sqrt{2}\over \sqrt{3}}\right)\,.
\ee

We find that the matrix $x_{IJ}$~\eqref{Xij} has a zero eigenvector and the corresponding eigenvector reads
\begin{equation}
\begin{split}
\xi_0=\left(\frac{4-D}{2},\frac{D-4}{2},1\right).
\end{split}
\end{equation}
So we should have the relation~\eqref{zeroeigen}.  On the other hand, since $\vec \d=\vec\g_2$, as we argued in the previous appendix $\theta=2$. However, a solution does not exists because other equations are not compatible with this value. We check directly that for the vectors given in~\eqref{2vectors}, the equations~\eqref{g20} do not have solution for three charges $(Q_1,Q_2,Q_3)$.

We still need to look at solutions where some of the vectors do not participate. These cases are as follows.

\begin{enumerate}

\item Only $\vec\gamma_1$ participates, others subleading.

In this case we set $Q_2=Q_3=0$. We find from~\eqref{g19} that
\begin{equation}
\vec\kappa=\left(\frac{\theta+2 z-4}{\sqrt{3}},\sqrt{\frac{2}{3}}(z+1-\theta)\right),
\end{equation}
and then
\begin{equation}
\vec\kappa\cdot \vec\delta=\theta-2\,.
\end{equation}
It is not consistent with the second condition of~\eqref{g17}, i.e., $\vec\kappa\cdot \vec\delta=\theta$. So the $Q_1$-charge solution does not exist.

\item Only $\vec\gamma_2$ participates, others subleading.

We set $Q_1=Q_3=0$. We obtain from~\eqref{g19} that
\begin{equation}
\vec\kappa=\left(\frac{\theta-2}{\sqrt{3}},-\sqrt{\frac{2}{3}}(\theta-2)\right),
\end{equation}
and then
\begin{equation}
\vec\kappa\cdot \vec\delta=\theta-2\,.
\end{equation}
The above relation is contrast to the second condition of~\eqref{g17}, which means that the $Q_2$-charge solution does not exist.

\item Only $\vec\gamma_3$ participates, others subleading.

We consider the case $Q_1=Q_2=0$, then
\begin{equation}
\vec\kappa=\left(\frac{3+D(z-1)-5 z+\theta}{\sqrt{3}},\frac{D(z-1)-2(\theta+z-3)}{\sqrt{6}}\right).
\end{equation}
$(\theta,z)$ can be solved from~\eqref{g17}, which are given by
\begin{equation}
\theta=D^2-8 D+20\sp z=-1\,.
\end{equation}
We will discuss this kind of solution in more detail later.

\item $\vec\gamma_1$ and $\vec\gamma_2$ participate the third is subleading.

After setting $Q_3$=0, we find that in order to satisfy $\kappa\cdot \vec\gamma_1=\kappa\cdot \vec\gamma_2$~\eqref{g17}, one should demand
\begin{equation}
Q_1^2(2+z-\theta)=0\,.
\end{equation}
Since $2+z-\theta\neq 0$ as otherwise  we obtain a logarithmically running gauge field, we have to further choose $Q_1=0$, which reduces to case 2 and there is no consistent solution.

\item $\vec\gamma_1$ and $\vec\gamma_3$ participate, the third is subleading.

We choose $Q_2=0$ and we are left a theory with two gauge fields.  Similarly, we can compute
\begin{equation}
x_{IJ}=\left(\begin{matrix} 3, & D-4\\
D-4, & {(D-4)^2\over 2}
\end{matrix}\right),
\end{equation}
 \begin{equation}
 x_I=\left(1,  0\right)\sp x_0=1\,,
\end{equation}
from which we can find
\begin{equation}\label{gamma13}
\begin{split}
z=\frac{D^2-10 D+27}{D-5} \sp \theta=\frac{2(D-6)}{D-5}\,,\\
Q_1^2=\frac{2(D^2-9 D+22)}{D^2-10 D+29}\sp Q_3^2=\frac{4(5-D)}{D^2-10 D+29}\,.
\end{split}
\end{equation}
Since $D\geqslant 5$, there is no way to make $Q_3^2>0$. So we can not find consistent solutions in this case.

\item $\vec\gamma_2$ and $\vec\gamma_3$ participate the third is subleading.

We consider the case with $Q_1=0$ and try to solve the equations~\eqref{g17}-\eqref{g19}. The solution that satisfies all equations does not exist.

\end{enumerate}

\subsection{DBI action from string setup}
From above analysis, we find that one can obtain consistent hyperscaling-violating geometry~\eqref{hsv2scalars} by using the one charge string action in which only $\vec \gamma_3$ ($F_C$) participates. In this subsection we check what this gives for the DBI action.

Using the results of appendix~\ref{app2scalars}, we introduce the new basis,
\begin{equation}
\begin{split}
\vec e_1&=\left(\frac{(D-10) D+28}{\sqrt{3} \sqrt{((D-8) D+18) ((D-8)D+24)}},-\frac{\sqrt{\frac{2}{3}} ((D-7) D+16)}{\sqrt{((D-8) D+18) ((D-8) D+24)}}\right),\\
\vec e_2&=\left(\frac{\sqrt{\frac{2}{3}} ((D-7) D+16)}{\sqrt{((D-8) D+18) ((D-8) D+24)}},\frac{(D-10)D+28}{\sqrt{3} \sqrt{((D-8) D+18) ((D-8) D+24)}}\right),
\end{split}
\end{equation}
as well as
\begin{equation}
(\phi_1,\phi_2)=\Phi_1\,\vec e_1+\Phi_2\,\vec e_2 \sp \vec\Phi=(\Phi_1,\Phi_2)\,.
\end{equation}
Then we obtain a top-down string action related to our DBI scaling solutions.
\begin{equation}\label{closedstr}
S_{closed}=\int d^4x\sqrt{g_{E}}~\left[{\cal R}_E-{1\over 2}(\pa\vec \Phi)^2+V_0~e^{-\vec \Delta\cdot \vec \Phi}-{1\over 4}~e^{\vec \Gamma\cdot \vec \Phi}F_C^2\right],
\end{equation}
\begin{equation}
S_{D}=\int d^{4}x~e^{\vec \alpha\cdot\vec\Phi}~\sqrt{\det(g_{E}+e^{\vec\beta\cdot\vec\Phi}~F)}\,,
\end{equation}
with
\begin{equation}\label{solution}
\begin{split}
\vec\Delta&=\left(\frac{(D-8) D+20}{\sqrt{((D-8) D+18) ((D-8) D+24)}},\frac{\sqrt{2}(D-4)}{\sqrt{((D-8) D+18) ((D-8) D+24)}}\right),\\
\vec\Gamma&=\left(-\frac{(D-4)^2}{\sqrt{((D-8) D+18) ((D-8) D+24)}},\frac{(D-4) ((D-8) D+20)}{\sqrt{2} \sqrt{((D-8) D+18) ((D-8) D+24)}}\right),\\
\vec\alpha&=\left(-\frac{D (D+p-11)-4 p+34}{\sqrt{((D-8) D+18) ((D-8) D+24)}},-\frac{D^3-2 D^2 (p+3)+2
   D (8 p+5)-40 p+16}{2 \sqrt{2} \sqrt{((D-8) D+18) ((D-8) D+24)}}\right),\\
\vec\beta&=\left(\frac{(D-8) D+20}{\sqrt{((D-8) D+18) ((D-8) D+24)}},\frac{\sqrt{2}(D-4)}{\sqrt{((D-8) D+18) ((D-8) D+24)}}\right).
\end{split}
\end{equation}
The background solutions supported by the closed string action~\eqref{closedstr} are identical to EMD in~\eqref{emdaction} with the identification
\begin{equation}
\Phi_1\rightarrow\phi \sp \Phi_2\rightarrow 0 \sp F_C\rightarrow H\,,
\end{equation}
and the corresponding exponents from this string setup are
\begin{equation}
\theta=D^2-8 D+20\sp z=-1\,,
\end{equation}
with
\begin{equation}
L^2=\frac{(D^2-8 D+19)(D^2-8 D+20)}{V_0}\sp Q_3^2=\frac{4}{D^2-8D+19}\,.
\end{equation}
We notice that the above values of $(\theta,z)$ satisfy the Gubser bound as $D\geqslant5$.
The profile for $\Phi_1$ is given by
\begin{equation}
e^{\Phi_1}=\left(\frac{r}{\ell}\right)^{\sqrt{(\theta-2)(\theta-2z+2)}}=\left(\frac{r}{\ell}\right)^{\sqrt{((D-8) D+18) ((D-8) D+24)}}.
\end{equation}

Notice that in the present case the  two coupling functions in DBI action read
\begin{equation}
Z_1=e^{-\frac{D (D+p-11)-4 p+34}{\sqrt{((D-8) D+18) ((D-8) D+24)}} \Phi_1}, \quad Z_2=e^{\frac{(D-8) D+20}{\sqrt{((D-8) D+18) ((D-8) D+24)}} \Phi_1}.
\end{equation}
Using the formulae~\eqref{cc1} and~\eqref{cc2}, we obtain the magneto-resistance as well as Hall resistivity
\begin{equation}
\rho_{xx}\sim T^{\lambda_1}{\sqrt{{\t^2}+c_1 ~T^{\lambda_2}+c_2 ~h^2~ T^{\lambda_3}}\over \t^2+c_1 ~T^{\lambda_2}}\,,
\end{equation}
\begin{equation}
\rho_{xy}\sim {h\t \over \t^2+c_1 ~T^{\lambda_2}}\,,
\end{equation}
with
\begin{equation}
\lambda_1=-2\sp \lambda_2=2(D-4)(D-p-1)\sp \lambda_3=2(D-4)(D-p-1)+4\,,
\end{equation}
with $c_1, c_2$ constants.
Here we have turned off the parity odd term, i.e., $S=0$.

In order to realise the magneto-resistance behavior~\eqref{linearTh} we should require $\lambda_1=-1$, while in our present case we find $\lambda_1=-2$. Therefore, we do not achieve the magneto-resistance behavior~\eqref{linearTh} from the string setup~\eqref{closedstr}-\eqref{solution}. A more complicated string setup would be possible to obtain~\eqref{linearTh}.


\end{document}